%% file: main.tex
\documentclass[sigconf]{acmart}
\usepackage{booktabs} % For formal tables
\usepackage{listings}
\usepackage{multirow}
\usepackage[table,xcdraw]{}
\usepackage{subfig}
\usepackage{parcolumns}
\usepackage[ruled]{algorithm2e} % For algorithms

\setcopyright{rightsretained}
\begin{document}

\title{Evaluation of Docker Containers for Scientific Workloads in the Cloud}
%\titlenote{}
%\subtitle{Student Technical paper}
%\subtitlenote{The full version of the author's guide is available as
%\texttt{acmart.pdf} document}

\author{Pankaj Saha}
%\authornote{}
\affiliation{%
  \institution{State University of New York (SUNY) at Binghamton}
  \streetaddress{4400 Vestal Pkwy E}
  \city{Binghamton}
  \state{New York}
  \postcode{13902}
}
\email{psaha4@binghamton.edu}

\author{Angel Beltre}
%\authornote{}
\affiliation{%
   \institution{State University of New York (SUNY) at Binghamton}
  \streetaddress{4400 Vestal Pkwy E}
  \city{Binghamton}
  \state{New York}
  \postcode{13902}
}
\email{abeltre1@binghamton.edu}

\author{Piotr Uminski}
%\authornote{}
\affiliation{%
  \institution{Intel Technology Poland}
  \streetaddress{Slowackiego 173}
  \city{Gdansk}
  \state{Poland}
  \postcode{80-298}
}
\email{piotr.uminski@intel.com}

\author{Madhusudhan Govindaraju}
%\authornote{}
\affiliation{%
   \institution{State University of New York (SUNY) at Binghamton}
  \streetaddress{4400 Vestal Pkwy E}
  \city{Binghamton}
  \state{New York}
  \postcode{13902}
}
\email{mgovinda@binghamton.edu}
% The default list of authors is too long for headers.
\renewcommand{\shortauthors}{P. Saha et al.}

\begin{abstract}
The HPC community is actively researching and evaluating tools to support execution of scientific applications in  cloud-based environments. Among the various technologies, containers have recently gained importance as they have significantly better performance compared to full-scale virtualization, support for microservices and DevOps, and work seamlessly with workflow and orchestration tools. Docker is currently the leader in containerization technology because it offers low overhead, flexibility, portability of applications, and reproducibility. Singularity is another container solution that is of interest as it is designed specifically for scientific applications. It is important to conduct performance and feature analysis of the container technologies to understand their applicability for each application and target execution environment. 

This paper presents a (1) performance evaluation of Docker and Singularity on bare metal nodes in the Chameleon cloud (2) mechanism by which Docker containers can be mapped with InfiniBand hardware with RDMA communication and (3) analysis of mapping elements of parallel workloads to the containers for optimal resource management with container-ready orchestration tools. 
%Our experiments were conducted on NSF funded clouds -- Jetstream and Chameleon. 
Our experiments are targeted toward application developers so that they can make informed decisions on choosing the container technologies and approaches that are suitable for their HPC workloads on cloud infrastructure. Our performance analysis shows that scientific workloads for both Docker and Singularity based containers can achieve near-native performance. 

Singularity is designed specifically for HPC workloads. However, Docker still has advantages over Singularity for use in clouds as it provides overlay networking and an intuitive way to run MPI applications with one container per rank for fine-grained resources allocation. Both Docker and Singularity make it possible to directly use the underlying network fabric from the containers for coarse-grained resource allocation.

% \footnote{
% This work was supported in part by NSF grant OAC-1740263.}
\end{abstract}

%
% The code below should be generated by the tool at
% http://dl.acm.org/ccs.cfm
% Please copy and paste the code instead of the example below.
%



\copyrightyear{2018} 
\acmYear{2018} 
\setcopyright{acmcopyright}
\acmConference[PEARC '18]{Practice and Experience in Advanced Research Computing}{July 22--26, 2018}{Pittsburgh, PA, USA}
%\acmBooktitle{PEARC '18: Practice and Experience in Advanced Research Computing, July 22--26, 2018, Pittsburgh, PA, USA}
\acmPrice{15.00}
\acmDOI{10.1145/3219104.3229280}
\acmISBN{978-1-4503-6446-1/18/07}

\maketitle

\input{1_introduction}
\input{2_background}
\input{4_experimentalSetup}
\input{5_evaluation}

\input{5_evaluation_1}
\input{6_related_work}

\input{7_conclusion}

\input{8_acknowledgements}

\bibliographystyle{ACM-Reference-Format}
%\bibliographystyle{acm}
%\bibliography{sample-bibliography}
\bibliography{Mendeley,bib-a}

\end{document}

%% file: 1_introduction.tex
\section{Introduction}
Containerization technology has gained significant traction in recent years. Containers are the appropriate tool for software development landscape that has adopted microservices and DevOps. Containers have several well known benefits for use in cloud environments: (1) support for a uniform environment for testing and deploying applications; (2) seamless and continuous updates of microservices; and (3) agility to support different languages and deployment platforms. 
%textcolor{blue}{[Fixme. How are containers different than VMs. Mention about the use of cgroups, etc. to implement containers.]} \textcolor{orange}{Need to add} 
%they have become the focal point of evaluation to identify bottlenecks for the adoption in High-Performance Computing (HPC) environments. Containers provide developers the facility to bring a custom-tailored runtime environment without taking into account the underlying host software stack.

When containers were initially adopted for micro-services based architectures in the industry, they did not attract the attention of HPC community wherein the focus is primarily on MPI applications for large parallel tasks. Containers were also initially shunned due to reports of possible root escalation vulnerabilities. However, the growing list of features along with the potential to provide high performance, along with portability and reproducibility from development to production environment, has made it critical to evaluate current container technologies for HPC applications in the cloud.

%In our findings, we were able to identify the reasons behind the lack of support of containers in the HPC community. Also, root escalation and performance overhead hinder the acceptance of containers in HPC.

%Docker ~\cite{Merkel1994LinuxJournal.},
%\textcolor{blue}{[Add this Docker reference the first time it is mentioned.]}

%In this paper, we are focused on the evaluation of two Container technologies: Docker and Singularity.

%\textcolor{blue}{[FIXME: need a few lines on Docker and its features.]} %\textcolor{orange}{REPLY: [moved the text to background section2]}
%\textcolor{blue}{[FIXME: need a few lines on Singularity and its distinguishing features.]} \textcolor{orange}{REPLY: [moved the text to background section2]}

%Other emerging container technologies of significance include 
%Shifter~\cite{Gerhardt2017Shifter:HPC} and and Charliecloud~\cite{Priedhorsky2017Charliecloud}.
%\textcolor{blue}{[FIXME: need a couple of lines on Shifter and CharlieCloud stating their differentiating features.]} 

%Singularity since it resolves the existing fundamental drawbacks presented by the different container technologies such as  

%However, many HPC centers still do not support Singularity as their primary choice of shipping HPC applications.  
Docker~\cite{DOCKER1} containers are known as lightweight Virtual Machines, which provide networking capabilities and dedicated computational resources. Singularity containers are designed to efficiently in conventional HPC environments. While Slurm~\cite{Yoo2003SLURM:Management} is the most widely used scheduler for HPC applications, the industry has successfully used open source technologies in the cloud such as Apache Mesos~\cite{Hindman2011Mesos:Center} and YARN~\cite{Vavilapalli2013ApacheYARN} for resource management, and container orchestrators like Kubernates~\cite{KUBERNETES1} and Docker Swarm~\cite{Naik2016BuildingClouds}. As there is support for Docker with these technologies, containerized HPC applications can also avail features such as container migration, resource fairness, and fault tolerance.

%\textcolor{blue}{[FIXME: why is Slurm mentioned here? Does Singularity have features to work with Slurm?]} \textcolor{orange}{REPLY: [we wanted to mention that the primary HPC scheduler is Slurm, whereas the advent of docker into HPC can introduce many cloud ready schedulers and container orchestrator to manage the resources along with other features like *container migration, *resource fairness, *fault tolerant etc.]}

%But, there is not enough support for cloud-based  schedulers 
Singularity is designed to use the underlying HPC runtime environment for executing MPI applications, whereas Docker is designed to isolate the runtime environment from the host. Also, Singularity focuses on coarse-grained resource allocation whereas Docker can take advantage of the fine-grained allocation of resources per rank.

HPC centers and academic clusters currently do not widely support Docker due to reports of security concerns that root escalation is possible. However, this vulnerability is not a concern in cloud allocations wherein users have root privileges to run their applications and other security modules provide separation between different allocations. As containers keep expanding support for running HPC tasks in the cloud, it is critical to quantify the impact on performance, support for orchestration and scheduling/placement of containers on the resources. 

The features of interest to the HPC community, which we present in this paper, include (1) evaluation of support for Infiniband and RDMA communication across MPI ranks; (2) mapping of sshd ports between the container and host machine; (3) determining overhead of containers compared to bare-metal access for memory, cpu, and communication intensive tasks; and (4) study use of overlay networks by container orchestration tools and its effect on HPC applications. 

Our experiments were conducted on an NSF funded academic Chameleon cloud~\cite{Mambretti2015NextSDN}. It provides a utility called "Complex Appliances," which offers accessible cluster configurations for the acquired nodes. One of the appliances is called "MPI bare-metal cluster," which comes with "MPICH2" library on CentOS 7 based image. Chameleon Cloud provides InfiniBand supported bare metal nodes.
%from Texas Advance Computing Centers (TACC). 

\begin{itemize}
	\item{We mapped InfiniBand devices of the host machine to Docker containers for RDMA communication across MPI ranks.} 
	\item{We quantified the performance overhead of MPI applications when run in containers and compared with their bare metal performance. For this evaluation, we used both the host and Docker's overlay network to execute the MPI benchmarks. With Docker, we orchestrated containers using different approaches:}
	\begin{enumerate}
			\item{We ran the container\textquotesingle s sshd on a non-standard port and mapped the port with host machines. One container per host was used in this approach such that the container and host node use the same IP address, but run sshd on different ports.}
			\item{We used an overlay network, provided by Docker Swarm, which spans across multiple host nodes. We created a {\it{one container per node}} setup and plugged all of them to the same overlay network. This network assigned separate IP addresses to all containers under the same subnet address.}
			\item{We used a similar setup as approach (2), but used multiple containers per host node attached to the common overlay network.}
	\end{enumerate}
    \item We used different classes (CPU , memory, and latency sensitive) of MPI applications for benchmarking. We observed for fixed number of MPI ranks how the performance varies when the number of nodes changes in the cluster. We also compared communication with InfiniBand vs Ethernet for the different classes of applications.
\end{itemize}

%% file: 2_background.tex
\section{Background}

Linux container (LXC) provides a virtual layer on top of the Linux host Operating System (OS) allowing multiple Linux systems to run in isolation. Containers separate the runtime environment from the underlying host resources and networking capabilities. Containers primarily rely on the use of namespace and cgroups to provide isolation, which enables users to run numerous applications in isolation on different containers on the same host. Additionally, due to the system level of abstraction, a container can freely move between host machines that support the container's runtime environment. Docker is the leading container solution widely accepted in industry. 
%Other container mechanisms in use today include FreeBSD jail and Solaris Zones. \textcolor{blue}{[FIXME: can't just mention these two other examples, and instead to list all the well known ones.]}
Docker has gained widespread acceptance in the recent years as can be seen by the support in resource managers and orchestration frameworks like Apache Mesos and Kubernetes. Another container technlogy that has gained the attention of the community is Singularity. It has been developed for scientific applications keeping the HPC eco-system as the focus. In the following subsections, we introduce these container mechanisms and their associated capabilities.

\subsection{Docker Container}
Docker can be considered a high-level user-space Linux utility, which can build, run and ship containers across hosts. Docker provides an isolated runtime environment. It is a lightweight Virtual Machine (VM) that has a networking configuration under a subnet and dedicated computational resources. A traditional VM abstracts the underlying hardware from guest OS, whereas Docker container provides one or more levels of abstraction by hiding the underlying host OS. Unlike other container solutions, Docker provides a virtual network, on top of the host machine, which can connect all containers to provide a convenient and secure inter-container communication.

\subsection{Docker Swarm Mode}
While containers provide a flexible packaging solution for building and shipping applications, additional tools are needed to manage the orchestration of multiple containers when running a distributed application. Docker Swarm addresses this need. It provides native support to manage the Docker container orchestration. This orchestration tool offers a software-defined overlay network across all participating host machines.  In this setup, containers reside under the same virtual network and communicate with each other without any other networking configurations. An application can run as part of a swarm service, which can be scaled up and down as required. Swarm also can provide an attachable overlay network where containers from any host can be connected during creation, and all attached containers can be part of the same network. 

\subsection{Singularity Containers}
Singularity has gained traction in the scientific community for its in-built support to integrate with the Message Passing Interface (MPI). Singularity provides an easy packaging mechanism for applications along with a user friendly runtime execution environment. Singularity is designed to execute containers like a normal process in the host machine. This feature makes the integration of Singularity with the HPC schedulers easy. A fundamental difference with Docker container is the file format Singularity uses to store the image. Unlike Docker, which uses multiple layers, Singularity stores the entire image as one large file. Singularity hub provides the repository to save and maintain public Singularity container images.

%% No need for this subsection for this paper
%\subsection{Message Passing Interface (MPI)}

%Scientific applications are predominantly based on MPI, which brings the power of distributed parallelism in the resource intensive tasks. OpenMP is a programming platform built to parallelize processes in a shared memory system over multiple processes running on a system. 
%Each MPI application runs across multiple processes (also called ranks), and each process can be distributed across several hosts. 
%In theory we can run as many MPI ranks as we like on a single node. However it does not make sense to run more ranks than physical resources that can be used by the application. 
%The optimal no of ranks to depends on the application. For some MPI applications, which internally use OpenMP to handle multiple threads can incorporate extra level of parallelism across ranks in a single node.  For very old, compute-intensive legacy applications created before threading become popular, each rank use a single thread so we should run as many ranks as the many cores present in the cluster. When application performs less computations and uses more memory accesses, we can use both threads per core. Using multiple threads only make sense when an MPI application uses OpenMP internally to take advantage of shared memory for ranks on the same node.

\subsection{InfiniBand} 
%There are a large number of applications that are communication inten exchange information through network communication for which an ideal case scenario will be to process packets with high throughput and low latency. 
InfiniBand is well known for providing high throughput and low latency communication in distributed and parallel applications. For simplification, InfiniBand uses two channel adapters (CA) (1) Host CA (HCA) and (2) Target (TCA). Among the two channel adapters, only HCA provides visibility of the underlying hardware and software necessary for communication. 
%Mellanox and Intel are the leading vendors of InfiniBand adapters and switches. 
The OpenFabrics Alliance controls the standard for developing software stack for the RDMA through InfiniBand. Vendors design and implement custom Verbs interfaces following the Verbs' specifications, which aims to address the interfaces that abstract the hardware components.   

%% file: 4_experimentalSetup.tex
\section{Experimental Setup}
We acquired six bare metal nodes from the Chameleon Cloud platform equipped with Mellanox InfiniBand interconnect. Each node consists of 48 cores and 128GB of RAM. Initially, each MPI benchmark was profiled to understand its characterization. For our experiments, a bare-metal execution is considered the baseline performance. We then use it to derive the respective performance variation for each containerization approach. Our set of benchmarks is composed of HPCG and miniFE for measuring the computational work, OSU-Micro benchmarks for measuring latency, and KMI Hash to profile memory bound applications. Each benchmark is profiled using experimental setups described in Table~\ref{table:orchestrations}.
%to measure the relative performance deviation of each container configuration to bare metal performance. 
%Every experiment was conducted under the same hardware specifications and software stack. 
We conducted ten iterations of each experiment and report the average.

\begin{table}[h!]
\centering
\begin{tabular}{|l|p{0.60\linewidth}|} 
 \hline
 {\bf Benchmarks} & {\bf Description}\\
 \hline
 HPCG 		& High Performance Conjugate Gradient \\ 
 \hline
 MiniFE 		& Unstructured finite element solver\\  
 \hline
 OSU	& Latency over MPI ranks\\ 
 \hline
KMI Hash	& Memory intensive integer operation\\ 
 \hline
\end{tabular}
\smallskip
\caption{\textit{Software Stack and Version}}
\vspace{-1.5em}
\label{table:benchmarks}
\end{table}
\begin{table}[h!]
\centering
\begin{tabular}{|l|p{0.60\linewidth}|} 
 \hline
 {\bf Software} & {\bf Version}\\
 \hline
 Linux		& CentOS 7.4.1708 \\ 
 \hline
 Open MPI 		& Open MPI-3.0.0 \\  
 \hline
 Infiniband		& ConnectX3  \\ 
 \hline
 Drivers 		& Mellanox OpenFabrics Enterprise Distribution  \\
 \hline
\end{tabular}
\smallskip
\label{table:softwareStack}
\caption{\textit{Software Stack and Hardware Components}}
\vspace{-1.5em}
\end{table} 
\begin{table*}[t]
\centering
%\caption{Container Orchestration Methods with Docker}

\begin{tabular}{|l|l|l|l|l|l|}
\hline
{\it Orchestration Method} 						& {\it sshd Port}     	& {\it \# of Containers per Host} 	& {\it IP Address}        	& {\it Network for mpirun and ssh}         \\ \hline
Bare Metal  (Section \ref{BM})       			& default (22) 			& 0                     			& host IP address   		& host network    						\\ \hline
Docker: Host Network (Section \ref{OCHN})       & custom (9100) 		& 1                     			& host IP address   		& host network    						 \\ \hline
Docker: Overley Network-I (Section \ref{OCON})  & default (22)  		& 1                     			& unique IP address 		& overlay network 						 \\ \hline
Docker: Overlay Network-II (Section \ref{MCON}) & default (22)  		& n $>$ 1                     		& unique Ip address 		& overlay network 						  \\ \hline
Singularity (Section \ref{SC})  		 		& default (22)  		& 1                    				& host IP address 			& host network	 					      \\ \hline
\end{tabular}
\vspace{1em}
\caption{Container Orchestration Methods}
\label{table:orchestrations}
\end{table*}

\subsection {Bare Metal Nodes + InfiniBand (IB)} \label{BM}
\textcolor{blue}{}
%We evaluated and profiled each benchmark on Bare-metal nodes via extensive execution of four sets of experiments  (1) HPCG with 72 ranks, (2) MiniFe with 96 ranks,(3) OSU with 96 ranks, and (4) KMI Hash with 72 ranks. We conducted the profiling on six {~\it CentOS 7} machines in which the essential InfiniBand interconnect drivers for RDMA and OpenMPI 3.0.0 were installed to enable MPI execution.

%In another approach with bare metal nodes, we have varied the size of the cluster growing from two nodes to six nodes while keeping the number of MPI ranks fixed for an application.
Cloud bare metal nodes with InfiniBand hardware were used int this setup. Open MPI 3.0.0 was configured with Mellanox interconnect driver, {\it ConnectX3}, for accessing InfiniBand hardware to facilitate RDMA communication. Each MPI benchmark in Table~\ref{table:benchmarks} was installed separately on each node. Experimental results for each benchmark obtained by this configuration is considered as the base, and performance deviation of each containerization approach.

%################ Required Figure ##################
% \begin{figure}[h!]
%  \vspace{-1.5em}
%   \includegraphics[width=0.5\textwidth]
%   {images/bareMetalNodes.pdf}
%   \caption{{ 
%   \textcolor{blue}{\it Bare metal nodes connected with fabric network and equipped with InfiniBand drivers for MPI RDMA communication. Standard ssh port is used for inter node ssh and mpirun communications.}
%   }}
%   \vspace{-0.5em}
%   \label{bareMetalNodes}
% \end{figure}

\subsection {{Docker: one container per node, host network + InfiniBand}
} \label{OCHN}

%\textcolor{blue}{OLD: }In Figure \textcolor{blue}{label:OneDCPerNodeHostNetwork},~we configured one Docker container per host machine.  Each container is setup to use the same network as the host machine network, but with a custom ssh port exposed by the container ssh-daemon to serve to the host machine as an entry point. The custom ssh-port setup is imperative to launch the application. However, the communication of MPI task is made possible through the mapping of the InfiniBand devices to the container. In this process, all drivers and plugins must be installed and configured inside the Docker container. 

\begin{figure}[h!]
  \includegraphics[width=0.5\textwidth]
  {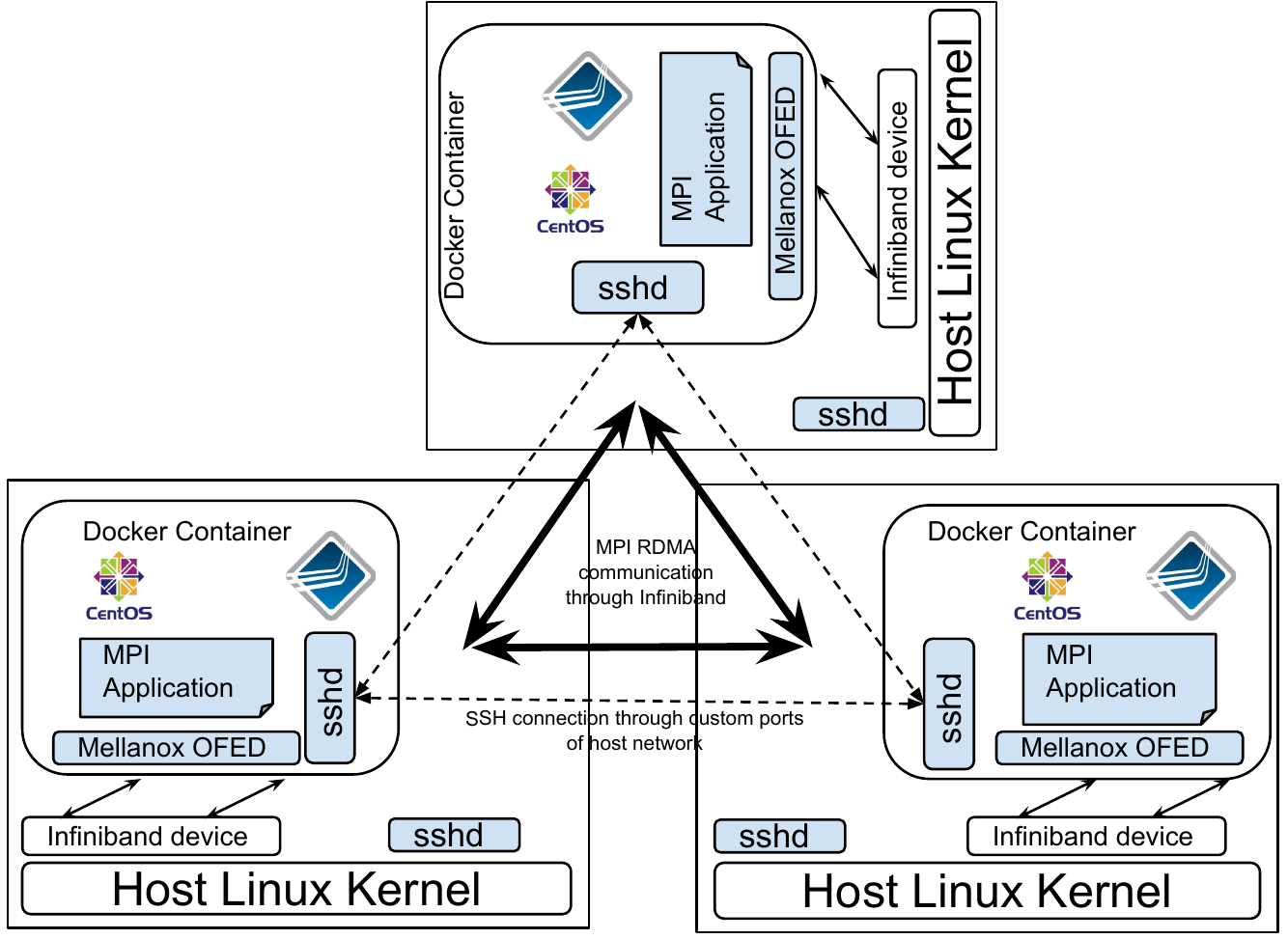}
  \caption{{ {\it One Docker container per host node is launched where each container can use the entire resource of the host node and host multiple MPI ranks. InfiniBand is used for MPI RDMA communication; however,  TCP/IP over Ethernet is used for ssh and mpirun by running sshd on container's nonstandard port mapped to host node.}
  }}
  \vspace{-0.5em}
  \label{OneDCPerNodeHostNetwork}
\end{figure}

Figure \ref{OneDCPerNodeHostNetwork} shows the required configuration for this setup. We configured one Docker container on each host node, where each container used the host network and shared the ip address of the host node. A new \textit{userid}, "mpi" was added to all the container images along with an ssh daemon configured to run on port 9100, so that it is different from the default port 22 used by the host's sshd daemon. For mpirun, hostfile contained the ip addresses of the participating host nodes and the host network was used. However, the RDMA communication of MPI ranks was made possible by direct mapping of the InfiniBand devices of host node to the container. At the time of container deployment, each container was mapped to the host with the required Mellanox InfiniBand devices. All the drivers were made available via the container by commands shown in Listing \ref{list:OCHN}. In order for the mapping of the devices to be accessible, containers needed to be started in privileged mode to access the devices of their host machine.

\begin{minipage}{0.45\textwidth}
\begin{lstlisting}[caption=Launching containers on different host nodes with custom sshd port mapped to host node,frame=tlrb,language=bash,label=list:OCHN] 
$ docker run --privileged 
-itd -p 9100:9100
--name host_container
--device=/dev/infiniband/rdma_cm 
--device=/dev/infiniband/uverbs0 
--device=/dev/infiniband/ucm0 
--device=/dev/infiniband/umad0 
--ulimit memlock=-1 
--network=host 
sciencecontainer/ompi3-infiniband:base
\end{lstlisting}
\end{minipage}
\vspace{-0.09in}

%################ Required Figure ##################

\subsection {Docker: one container per node, overlay network + InfiniBand}\label{OCON}

Docker swarm mode enables users to abstract the underlying host machine topology through a virtual overlay network. This network can be spun across a few or a large number of host machines. Even though the physical location of each container may be different, as they all may not share the same host machine, they all share the software-defined network (SDN) and subnet address. This setup enables direct communication across containers. Unlike the previous approach in section \ref{OCHN}, in this approach each container has a different IP address under the same subnet address. So, containers do not need to communicate over a non standard custom ssh port. It is important to note that we selected one of the containers as master-container and mapped its ssh port to the host node. The master container is used as an entry point to start the MPI application from within the container.
For mpirun, the {\it hostfile} contains the ip addresses of each container provided by the overlay network. RDMA communication across MPI rank was enabled as the host node's InfiniBand drivers were mapped as explained in the section \ref{OCHN}. 

\begin{minipage}{0.45\textwidth}
\begin{lstlisting}[caption=Creating custom overlay attachable network,frame=tlrb,language=bash]
$ docker network create 
--driver overlay --subnet 10.10.0.1/24 
--attachable custom-network
\end{lstlisting}
\end{minipage}

\begin{figure}[h!]
  \includegraphics[width=0.5\textwidth]
  {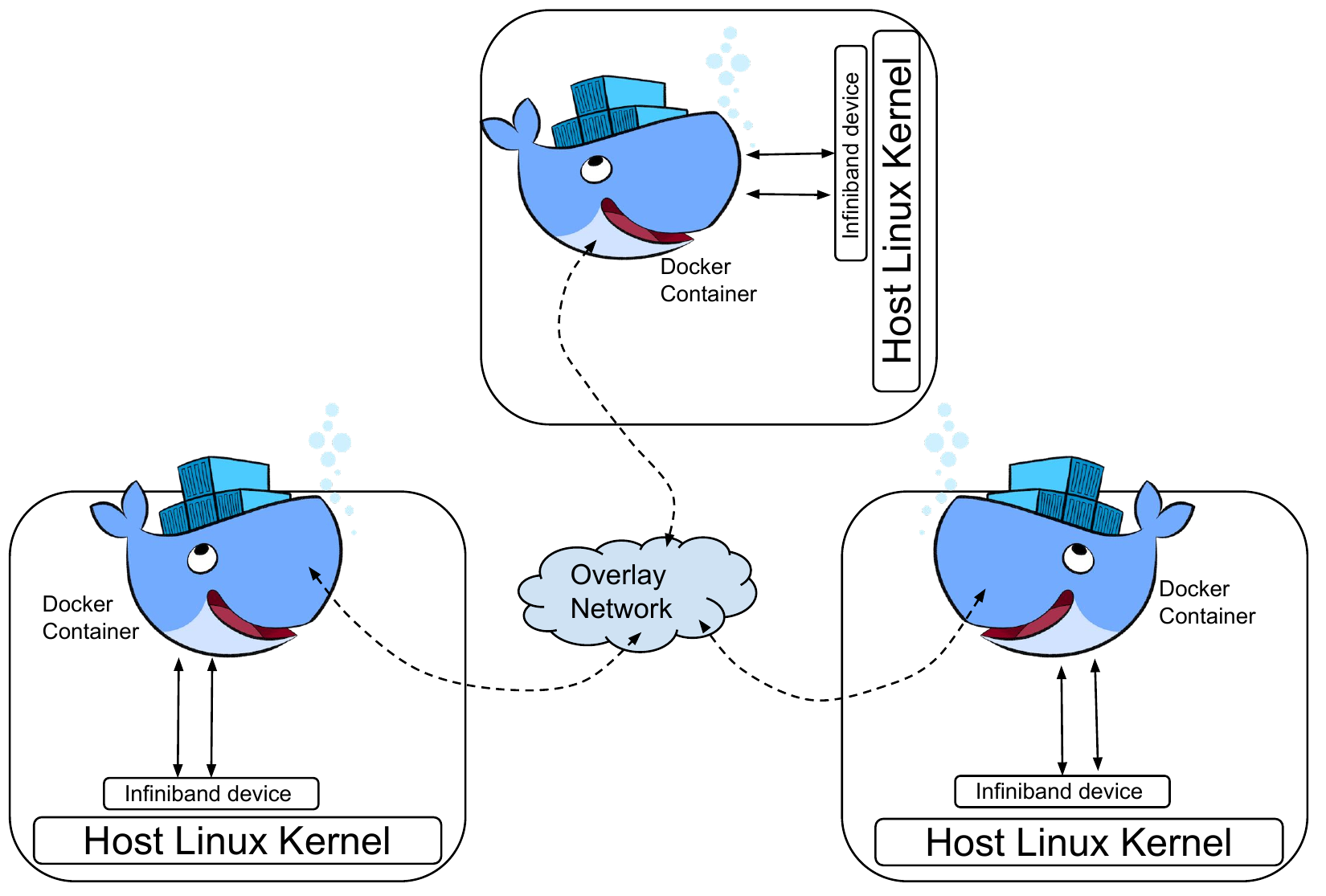}
  \caption{{{\it One Docker container per host node is launched where containers from different hosts are connected to each other through Docker Swarm defined overlay network. ssh and mpirun use overlay network over standard ssh port, whereas for MPI RDMA communication, InfiniBand is used.}
  }}
  \vspace{-0.5em}
  \label{OneDCPerNodeOverlayNetwork}
\end{figure}

\subsection {Docker - Multiple containers per Host and 'n' MPI Ranks per Container} \label{MCON}
In this approach we launched multiple containers per host node and we split the MPI ranks among containers across nodes. As described in Section~\ref{OCON}, each container is part of the Docker Swarm overlay network and for \textit{mpirun} the {\it hostfile} contains the ip addresses of the containers. A configuration similar to the one mentioned in Section \ref{OCON} was used while launching the containers via RDMA communication across MPI ranks and also for starting the MPI application.  

Initially we launched one container for one MPI rank and then started growing the number of ranks per container while keeping the total number of ranks same. Containers were equally divided among host nodes in the cluster and each container hosted equal number of MPI ranks. The performance difference in this setup is dependent on the number of ranks per container.

\begin{minipage}{0.45\textwidth}
\begin{lstlisting}[caption=creating Singularity images from Docker image ,frame=tlrb,language=bash]
$ sudo singularity build 
ompi3-infiniband-base.simg 
docker://sciencecontainer/ompi3:base
\end{lstlisting}
\end{minipage}\hfill

\begin{minipage}{0.45\textwidth}
\begin{lstlisting}[caption=Launching Singularity containers using host MPI libraries,frame=tlrb,language=bash]
$ mpirun -mca mtl_mxm_np 0 
--mca btl openib -np 72 
--map-by node 
--hostfile ~/hostfile 
singularity exec 
ompi3-infiniband-base.simg 
<path to executable inside container>
<parameters>
\end{lstlisting}
\end{minipage}
\subsection {Singularity Container} \label{SC}
We used MPI libraries of the host node to run applications through the Singularity containers. Running application through Singularity is like running the application on the host node. Singularity provides a way to create singularity images from existing Docker image layers. We used our Docker MPI images to create singularity images for the experiments. 

%% file: 5_evaluation.tex
\section{Evaluation }
\subsection{HPCG -High Performance Conjugate Gradients}

\begin{figure*}%
	\captionsetup[subfigure]{justification=centering}
	\centering
    \subfloat[{\it Performance comparison of HPCG benchmark with different approaches, running with 72 MPI ranks.} \label{fig:hpcg_approaches}  ]{{\includegraphics[width=0.30\linewidth]
    {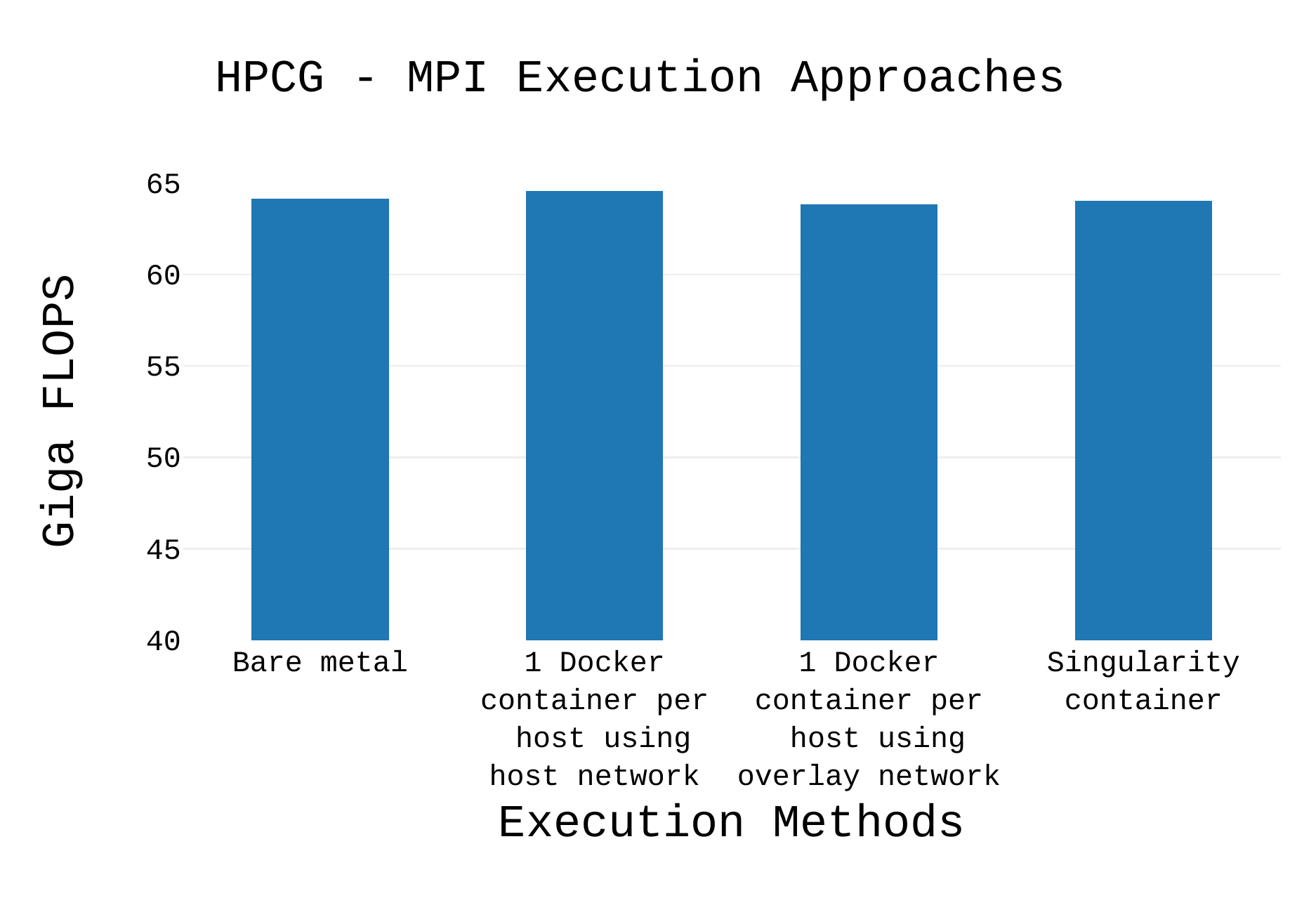}  
    }}
    \subfloat[{\it HPCG benchmark performance comparison when the benchmark is running across multiple Docker container per host node. A total of 72 MPI ranks were used and distributed equally in varied number of containers. Each node hosted equal number of containers.}]
{{\includegraphics[width=0.30\linewidth]
    {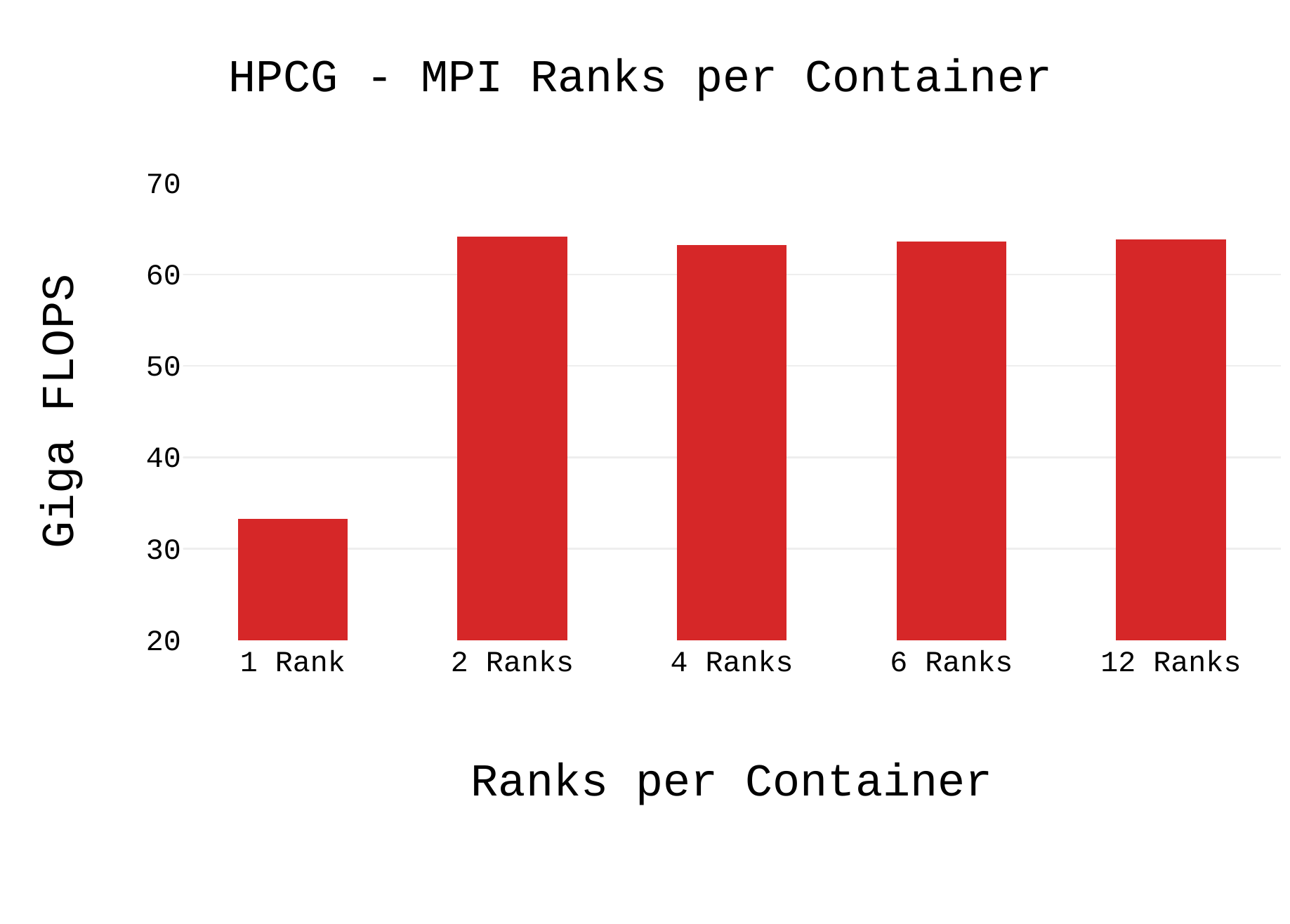}
    \label{fig:hpcg_ranks_per_container}
    }}
    \subfloat[{\it HPCG benchmark evaluation when 24 MPI ranks were distributed across variable number of hosts in a cluster.}]
{{\includegraphics[width=0.30\linewidth]
{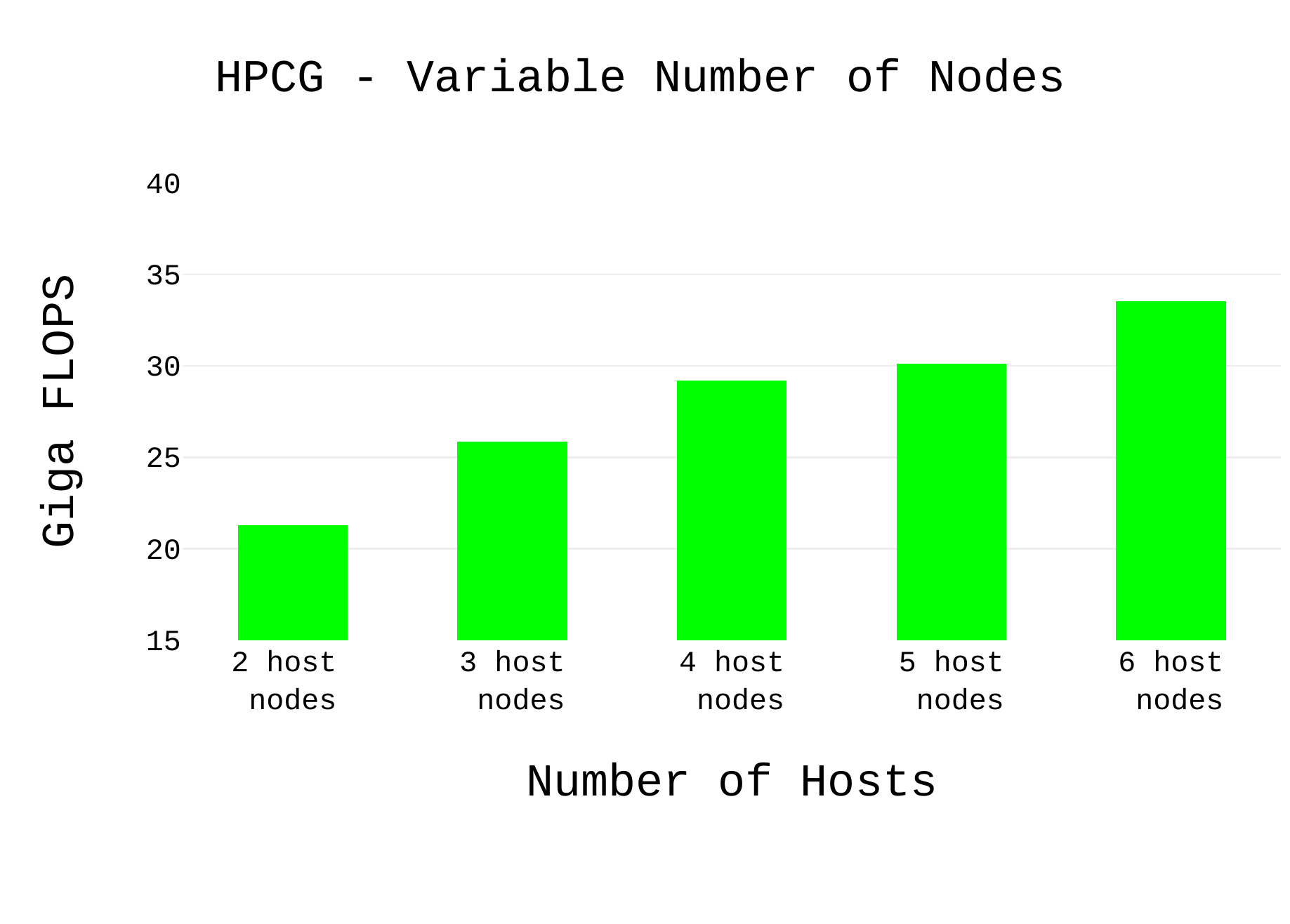}\label{fig:hpcg_multipleHosts}
    }}
\caption{HPCG Performance Evaluation with Different Execution Approaches.}
\end{figure*}

We ran the HPCG benchmark on our experimental cluster with 72 MPI ranks for all the approaches mentioned in Table~\ref{table:orchestrations}. In Figure \ref{fig:hpcg_approaches}, our results show that the relative performance overhead of different containerization approaches is small when compared to traditional bare metal approaches. The approach (Section \ref{OCHN}), { \it {One Docker container per host}} using host-network, yielded 0.65\% overhead compared to bare metal. However, { {\it one Docker container per host}} with overlay-network (section \ref{OCON}) and Singularity showed performance reduction by 0.47\% and 0.22\% respectively. 
%We ran 10 iterations of the same experiments and result varied  by less than 1\% in each iteration.

In Figure \ref{fig:hpcg_ranks_per_container}, we present HPCG performance when multiple MPI ranks are mapped with each Docker container. We ran { \it {one container per rank}} using 72 containers for a total 72 MPI ranks. We observed that the performance degraded by 50\% compared to bare metal. However, when the number of ranks per container is increased from one to an amount equal or greater than 2, we observed that performance was similar to bare metal performance. %\textcolor{blue}{[This is a bit vague. What is the threshold number of ranks after which the performance is the same.]} 

Figure \ref{fig:hpcg_multipleHosts} represents how the performance of the HPCG benchmark changes as we vary the number of nodes in a cluster. We varied the number of nodes from two (2) to six (6) and only executed 24 ranks throughout all the experiments.  Our experimental results showed that, as expected, performance increased as the number of hosts in the cluster increased. From a two-node cluster to a four-node cluster, the performance increased by 37\%. Additionally, after expanding the cluster by two more nodes, the performance increased by another 20\%. 

\subsection{MiniFE - Finite Element mini-application}
\begin{figure*}%
	\captionsetup[subfigure]{justification=centering}
	\centering
    \subfloat[\it Performance comparison of MiniFE benchmark with different approaches, running with 96 MPI ranks. \label{fig:minife_approaches}  ]{{\includegraphics[width=0.30\linewidth]
    {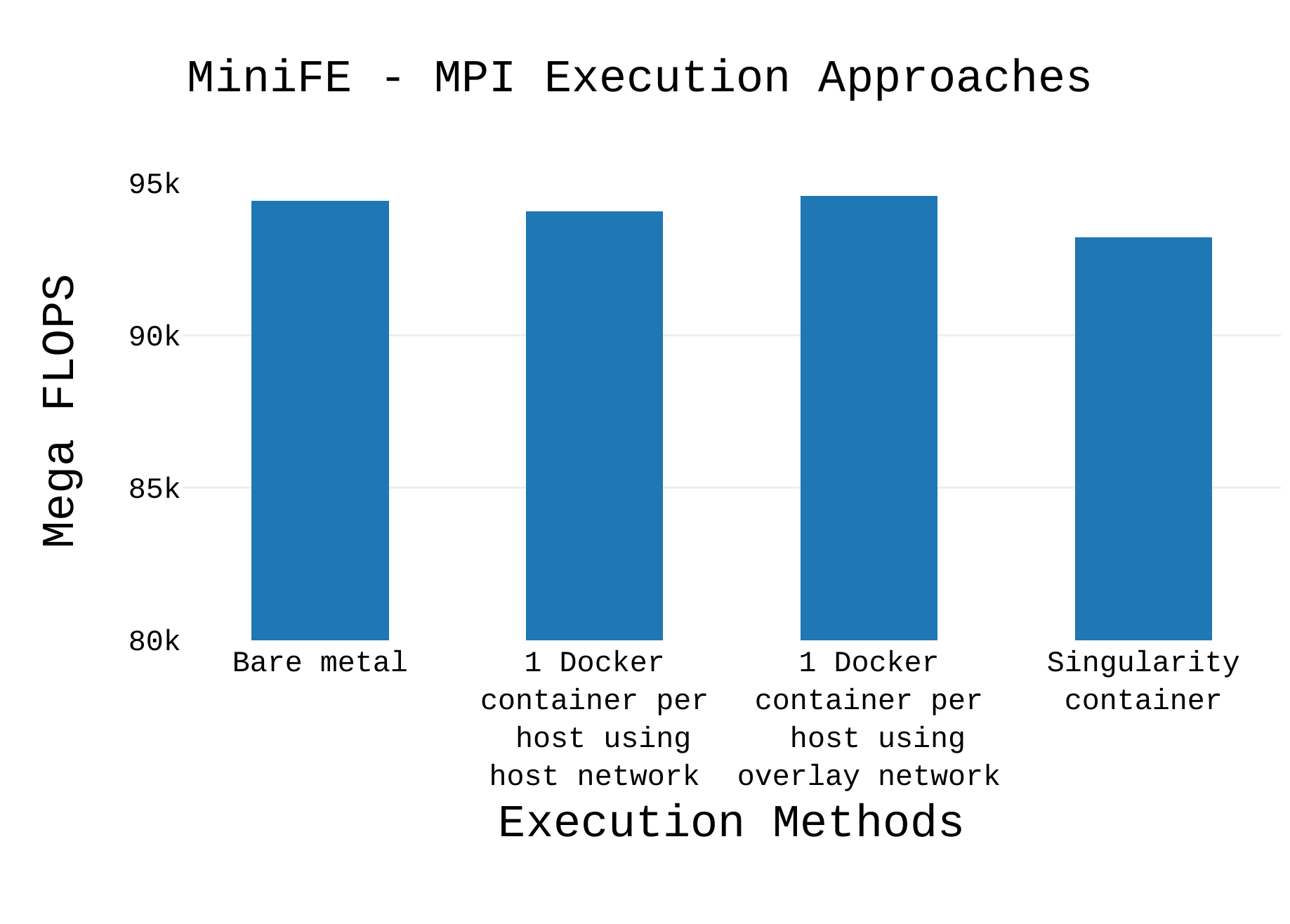}  
    }}
    \subfloat[\it MiniFE benchmark performance comparison when the benchmark is running across multiple docker containers per host node. A total of 96 MPI ranks were used and distributed equally in varied number of containers. Each node hosted equal number of containers.]
{{\includegraphics[width=0.30\linewidth]
    {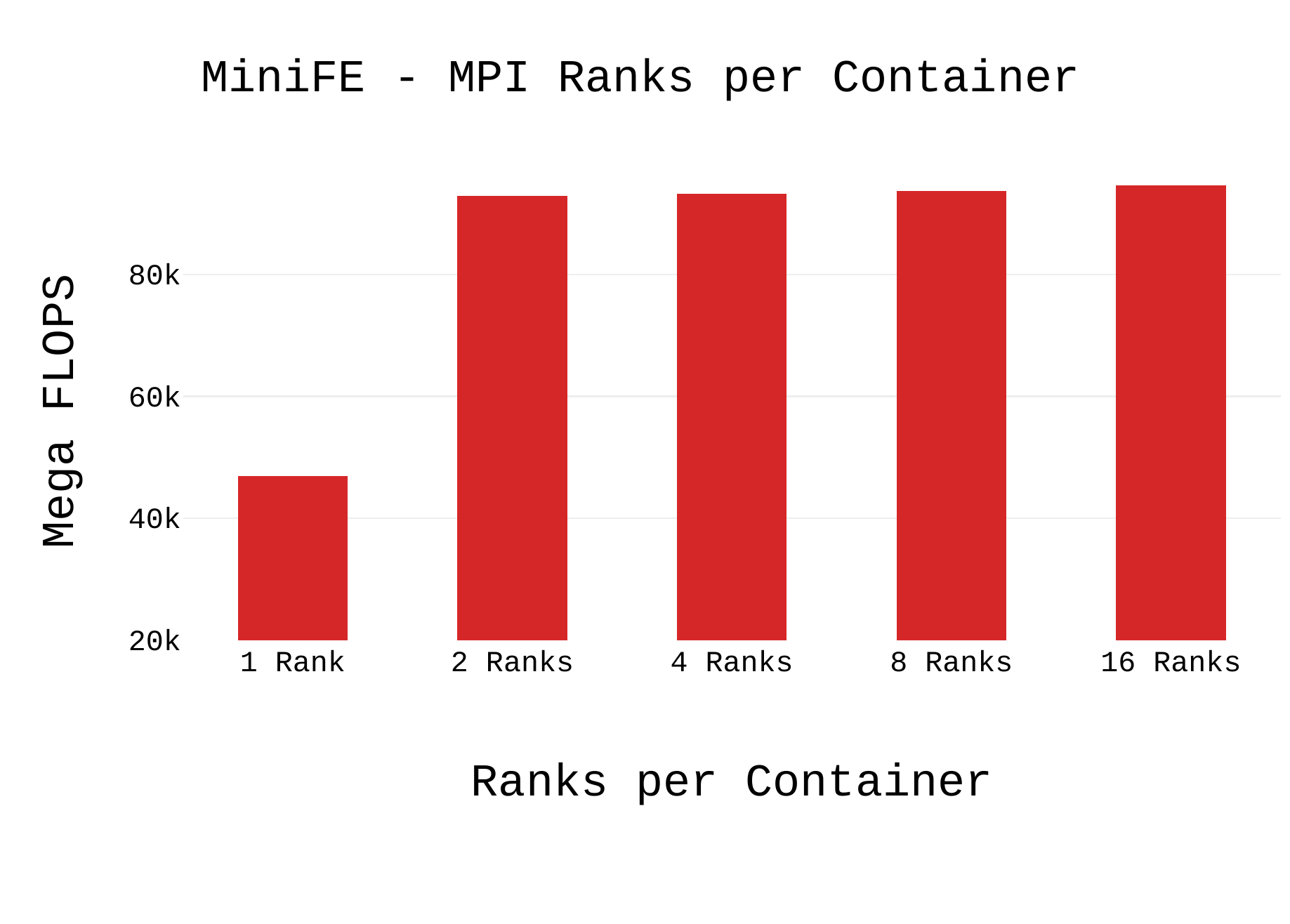}
    \label{fig:minife_ranks_per_container}
    }}
    \subfloat[\it MIniFE benchmark is evaluated when 32 MPI ranks were distributed across variable number of hosts in a cluster.]
{{\includegraphics[width=0.30\linewidth]
{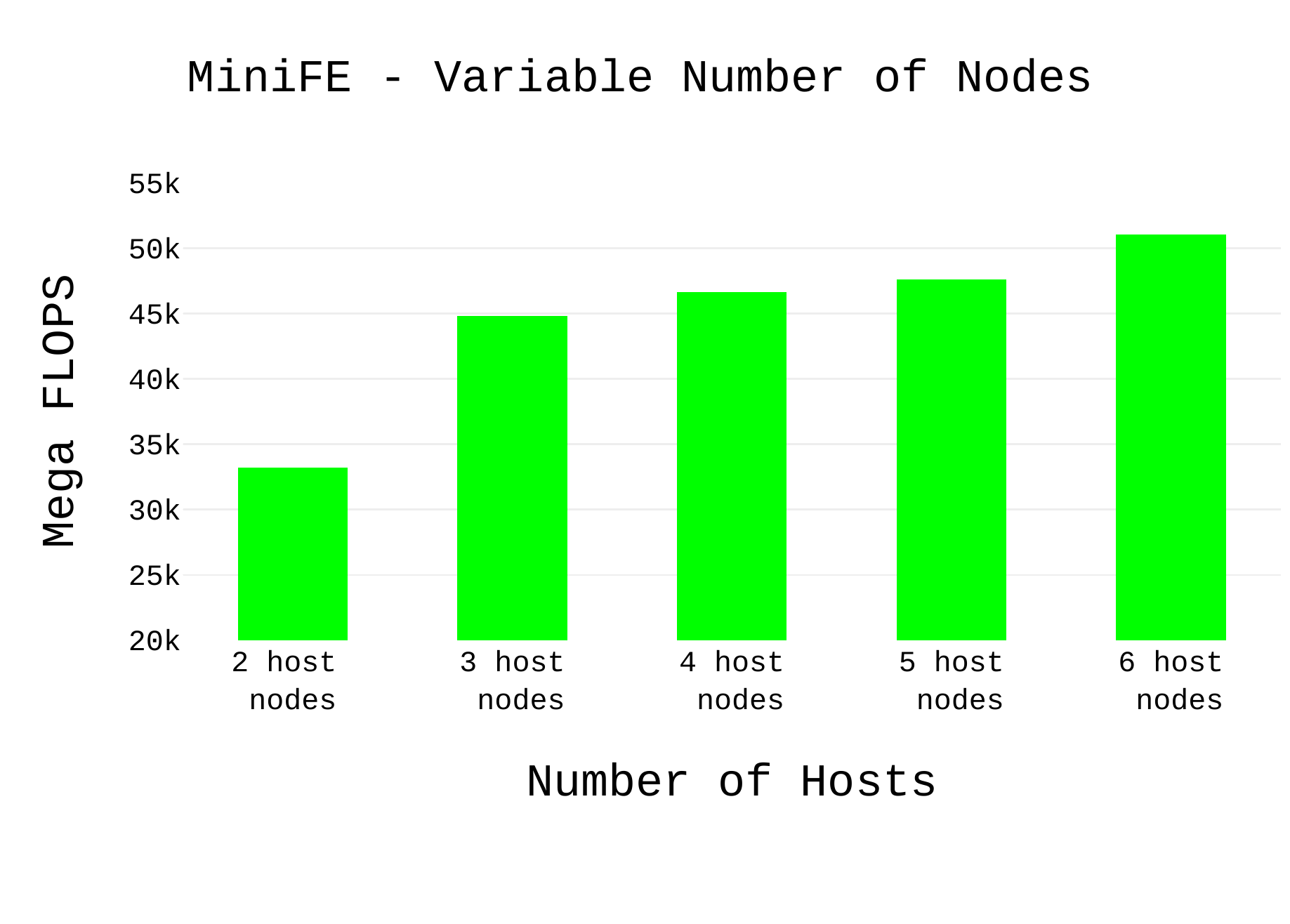}\label{fig:miniFE_multipleHosts}
    }}
\caption{MiniFE Performance Evaluation with Different Execution Approaches.}
\end{figure*}

Figure \ref{fig:minife_approaches} shows the performance of MiniFE benchmark with the various approaches when it ran with 96 MPI ranks. One container per host with the host-network yielded an overhead of 0.36\% compared to bare metal. One container per host with overlay-network produced 0.15\% degradation. On the other hand, when MiniFE ran with Singularity, we observed a 1.25\% overhead in performance. Like HPCG, we observed similar behavior for MiniFE, when ranks are divided into multiple containers.  

In Figure \ref{fig:minife_ranks_per_container}, { {\it one container per rank}} approach produced 50\% performance loss compared to bare metal, whereas when running two, four, eight, and sixteen ranks per container, the performance approached that of the bare metal setup.

We ran MiniFE with a fixed number of ranks (32 MPI ranks) while changing the size of the cluster from two nodes to 6 nodes. We observed that the performance expectedly improves as the number of nodes increases in the cluster. In Figure \ref{fig:miniFE_multipleHosts}, as the cluster size increased from two (2) nodes to four (4) nodes, the performance improved by 40\%. Also, adding another two nodes into the cluster improved the performance by another 13\% while keeping the number of ranks same. 

\subsection{OSU - Ohio State University Micro benchmarks}
% \textcolor{orange}{The OSU benchmark suite provides a set of mini-benchmarks, which allow profiling of network bandwidth performance as well as latency. To enable the collection of bandwidth related information, we ran \textbf{osu\_mbw\_mr} for which our results show that there is a performance hit in terms for 1, 4 and 8 ranks per container of \textcolor{green}{[okay]\%, [bad]\%, and [worst]\%} respectively. 
% For lantency profiling, we ran \textit{alltoallv} for which our results show that the latency did not change significantly for any of our approaches. When we split the ranks into containers, one container per rank performance is within 0.82\%  of the bare metal performance, whereas four (4) ranks per container yielded an overhead of 0.72\%. For the multi-node experiment, OSU ran with a fixed (24) number of ranks for an increasing number of hosts in which every variation outperformed the previous one -- starting from two (2) hosts to six (6) hosts, with an end latency improvement of 63\%.}

%\ref{fig:osu_approaches}
 We chose {\it alltoallv} collective communication benchmark from the OSU benchmark suite to measure the latency  across all the ranks when distributed across nodes. OSU can run  'n' processes where each sends 1/n of its allocated data to all the other ranks and receives a response back. We ran the latency benchmark for message size of 65536 Bytes. In Figure 5, our results show that the latency did not change significantly for any of our approaches. When we split the ranks into containers, { {\it one container per rank}} performance was within 0.82\% of the bare metal performance, whereas four (4) ranks per container yielded an overhead of 0.72\%. For the multi-node experiment, OSU ran with a fixed (32) number of ranks for an increasing number of hosts in which every variation outperformed the previous one -- starting from two (2) hosts to six (6) hosts, with a final latency improvement of 63\%. 

%We did not run the throughput experiments with the OSU suite since we expect the trends to be similar to the latency tests for these setups mentioned in Table \ref{table:orchestrations}. 

%\textcolor{blue}{[We earlier called OSU a communication intensive benchamrk, but talk about latency here. Needs to be corrected/fixed.]}

	%
    %
    %
    %
    %
    %
    %
    %
    %
    %
\begin{figure*}%
	\captionsetup[subfigure]{justification=centering}
	\centering
    \subfloat[\it Performance of OSU latency benchmark benchmark with different approaches, running with 96 MPI ranks \label{fig:osu_approaches}  .]{{\includegraphics[width=0.30\linewidth]
    {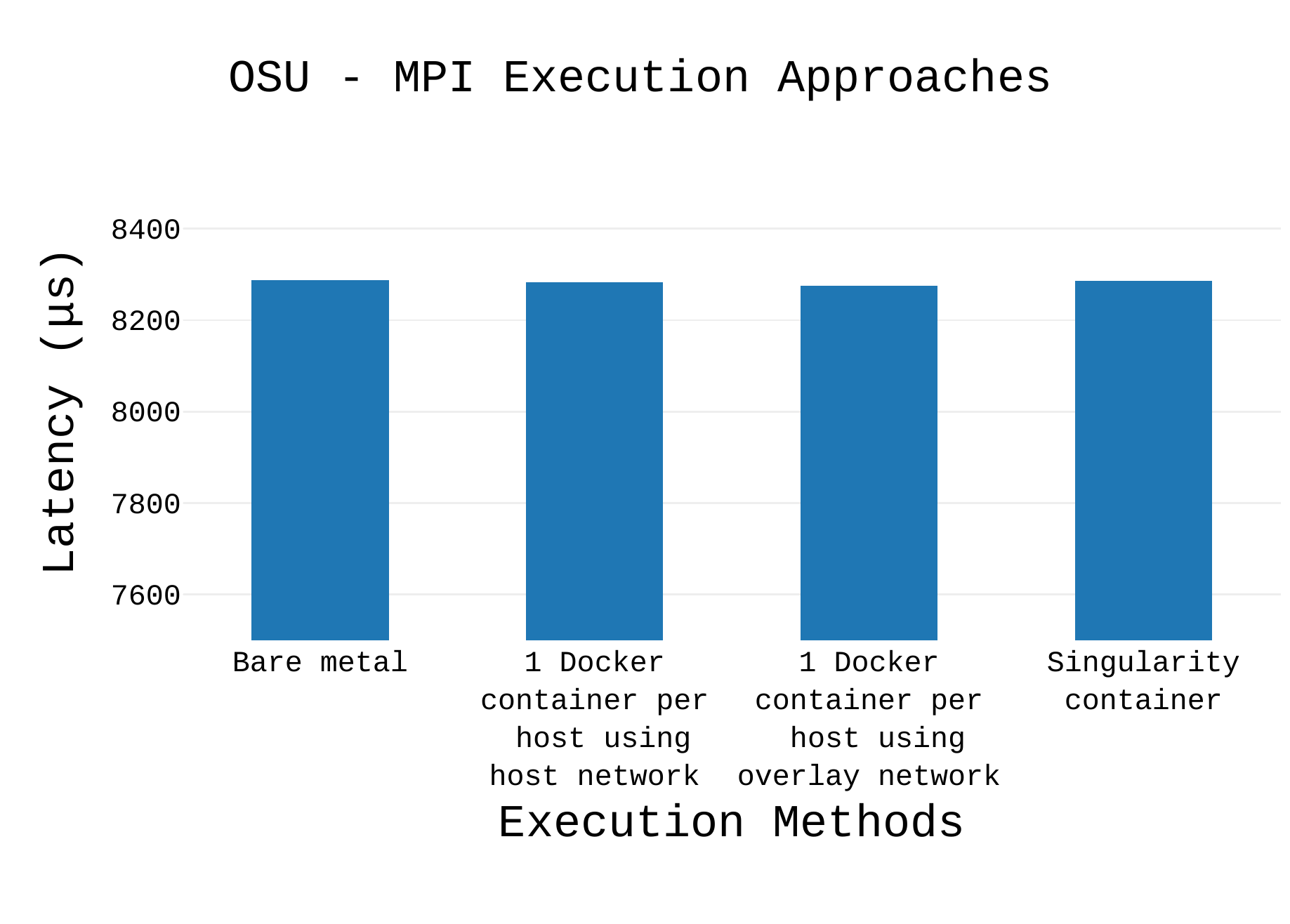}
    }}
    \subfloat[\it OSU latency benchmark performance comparison when the benchmark is running across multiple Docker container per host node. In total 96 MPI ranks were used and distributed equally in varied number of containers. Each node hosted equal number of containers.]
{{\includegraphics[width=0.30\linewidth]
    {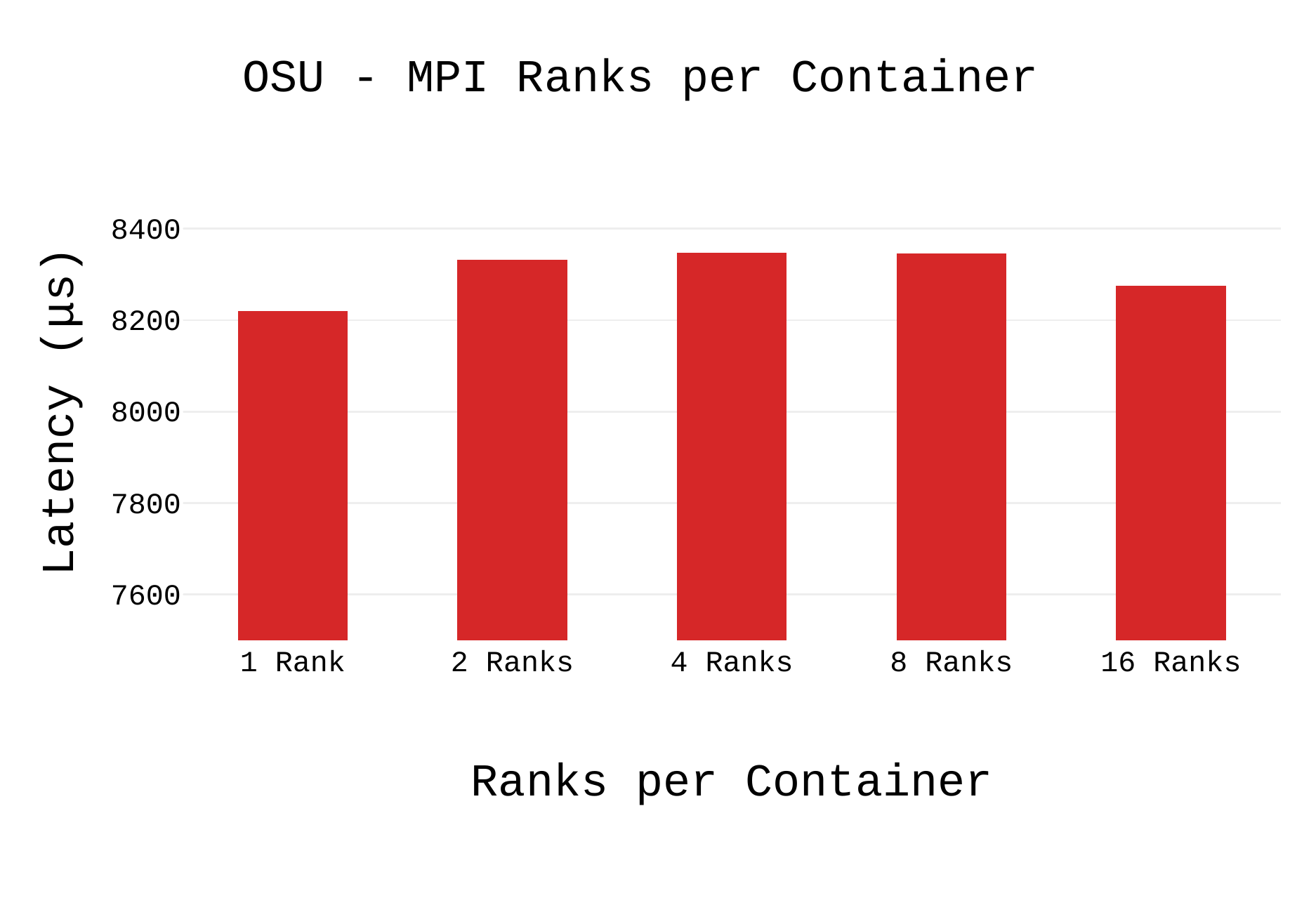}
    \label{fig:osu_ranks_per_container}
    }}
    \subfloat[\it OSU latency benchmark is evaluated with 32ranks, across variable number of hosts in the cluster.]
{{\includegraphics[width=0.30\linewidth]
{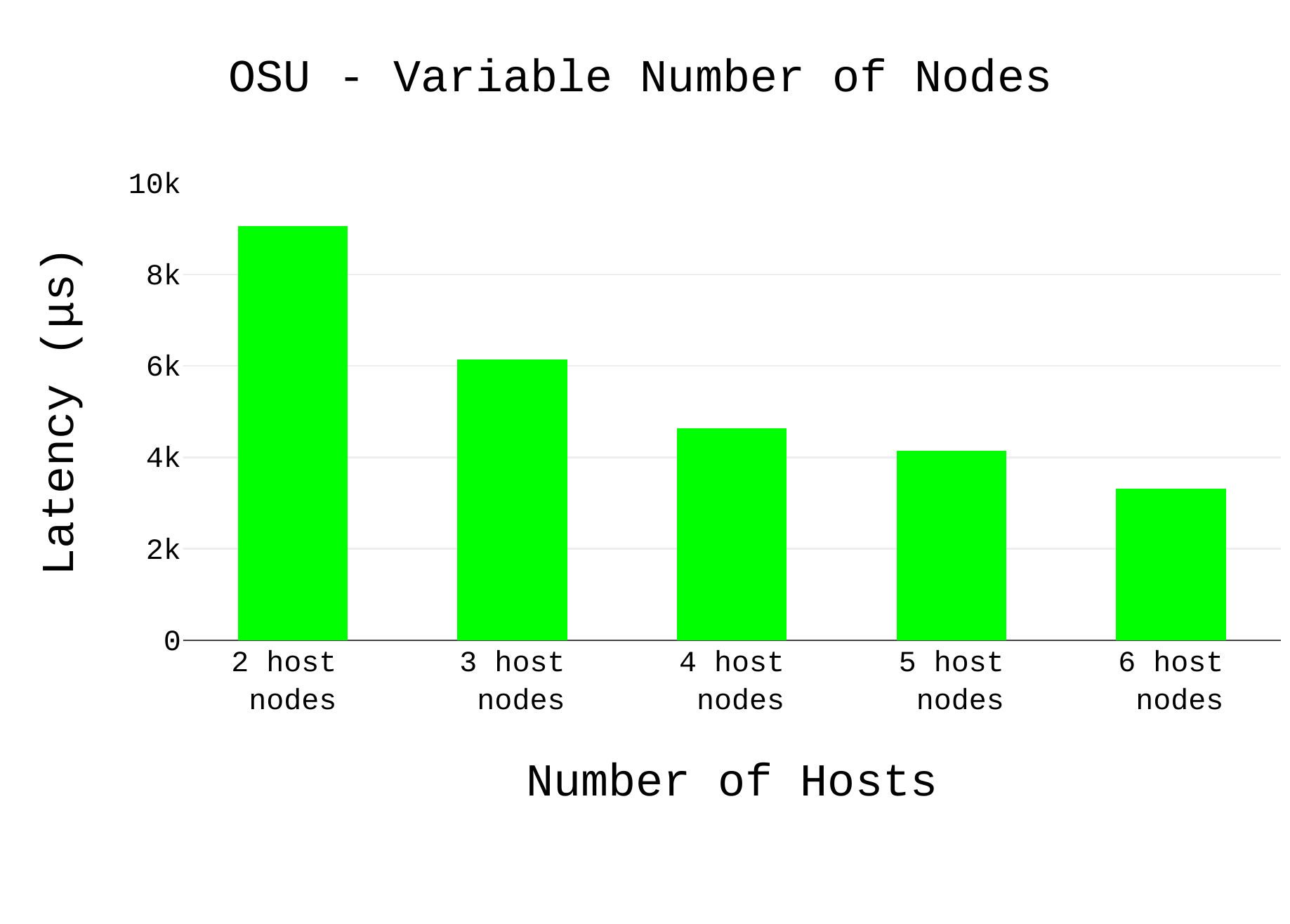}\label{fig:OSU_multipleHosts}
    }}
\caption{OSU (Latency) Performance Evaluation in Different Execution Approaches}
\vspace{-1em}
\end{figure*}
	%
    %
    %
    %
    %
    %
    %
    %
    %
    %
% \begin{figure*}%
% 	\captionsetup[subfigure]{justification=centering}
% 	\centering
%     \subfloat[\it Performance of OSU Bandwidth benchmark benchmark with different approaches, running with 96 MPI ranks \label{fig:osu_bw_approaches}.]{{\includegraphics[width=0.30\linewidth]
%     {images/osu_bw_approaches.pdf}
%     }}
%     %
%     %
%     \subfloat[\it OSU bandwidth benchmark performance comparison when the benchmark is running across multiple docker container per host node. In total 96 MPI ranks were used and distributed equally in varied number of containers. Each node hosted equal number of containers.]
% {{\includegraphics[width=0.30\linewidth]
%     {images/osu_bw_ranks_per_container.pdf}
%     \label{fig:osu_bw_ranks_per_container}
%     }}
%     %
%     %
%     \subfloat[\it OSU bandwidth benchmark is evaluated when 32 MPI ranks were distributed across variable number of hosts in the cluster.]
% {{\includegraphics[width=0.30\linewidth]
% {images/osu_bw_multiple_hosts.pdf}\label{fig:osu_bw_multipleHosts}
%     }}
% \caption{OSU Bandwidth Performance Evaluation in Different Execution Approaches}
% \vspace{-1em}
% \end{figure*}

\subsection{KMI Hash - M-mer Matching Interface benchmark}
\begin{figure*}%
	\captionsetup[subfigure]{justification=centering}
	\centering
    \subfloat[\it KMI Hash benchmark benchmark with different approaches  to compare the relative throughput for 72 MPI ranks.   ]{{\includegraphics[width=0.30\linewidth]
    {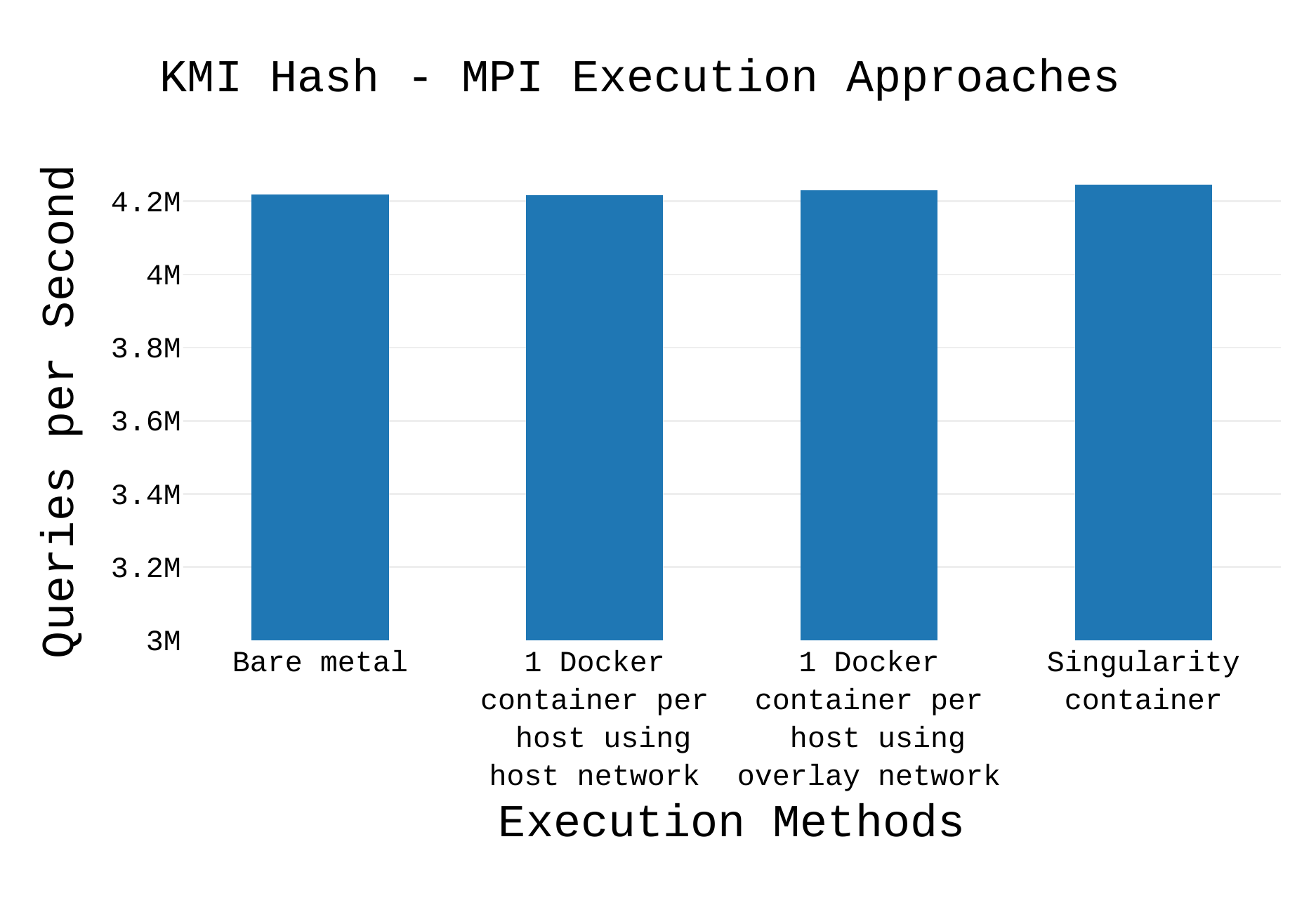}
    \label{fig:kmihash_approaches}  
    }}
    \subfloat[\it KMI Hash benchmark performance with different container to MPI rank ratio. In total, 72 MPI ranks were used and distributed equally in varied number of containers. Each node hosted equal number of containers.]
{{\includegraphics[width=0.30\linewidth]
    {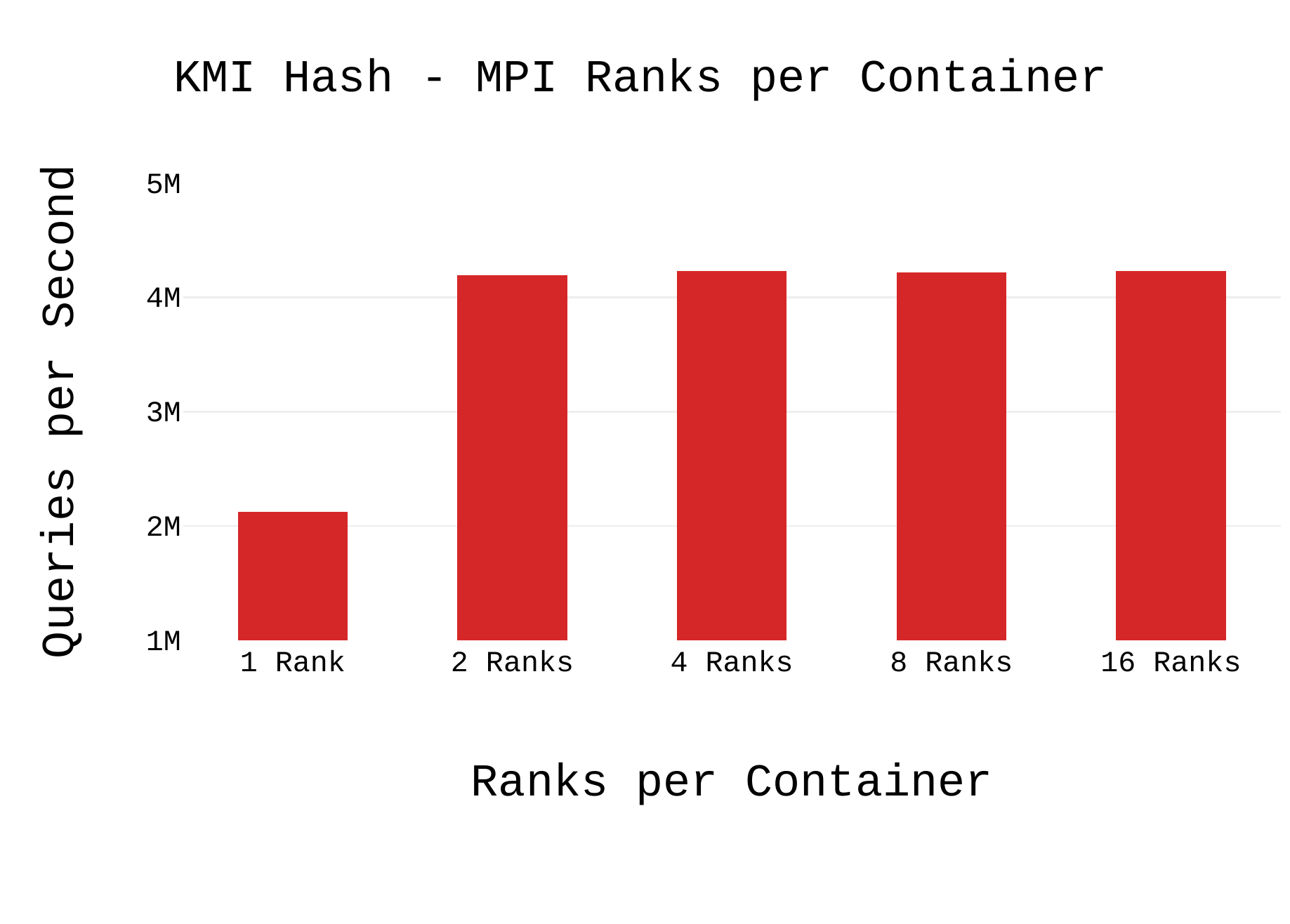}
    \label{fig:kmihash_ranks_per_container}
    }}
    \subfloat[\it KMI Hash - Performance Evaluation with 24 MPI ranks on different number of hosts in the cluster.]
{{\includegraphics[width=0.30\linewidth]
{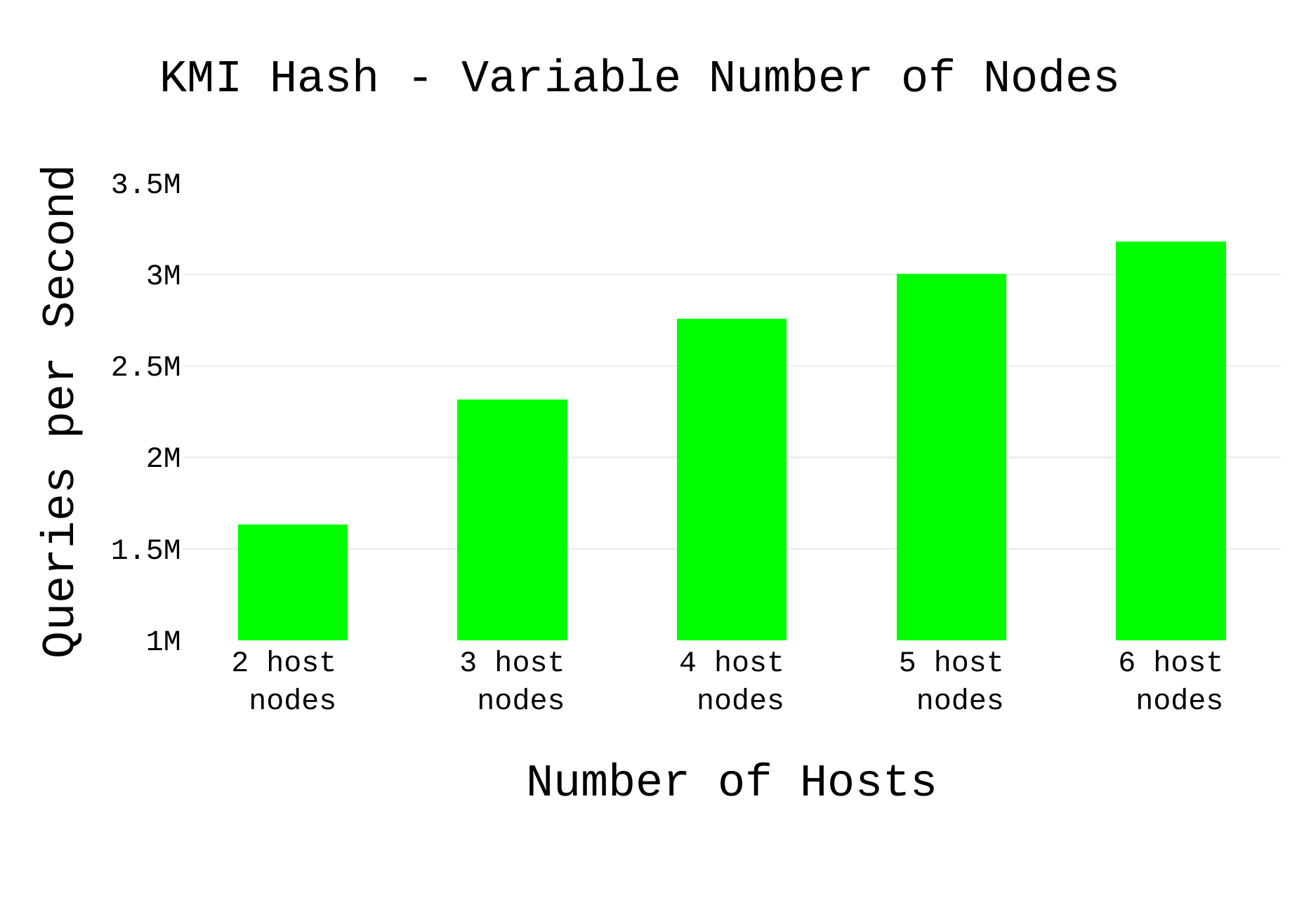}
\label{fig:kmihash_multipleHosts}
    }}
\caption{KMI Performance Evaluation with Different Execution Approaches.}
\end{figure*}

The purpose of KMI-Hash data-centric benchmark is to measure the performance of integer operation such as hashing. In Figure \ref{fig:kmihash_approaches}, our experimental results show that Docker container approaches produce similar throughput compare to the setup with bare metal nodes. Singularity performs within 0.6\% of the bare metal performance. In Figure \ref{fig:kmihash_ranks_per_container} apart from { {\it one container per rank}}, other approaches did not show any significant overhead compared to the bare metal setup. In Figure \ref{fig:kmihash_multipleHosts}, multi-node execution for KMI hash yielded similar trends in results as the previous benchmarks. However, for KMI Hash the performance improved by 94\%  as the number of hosts increased from two nodes to six nodes while the number of ranks was fixed to 32. 

%% file: 5_evaluation_1.tex
\subsection{Evaluating InfiniBand and Ethernet for different classes of MPI applications}
We conducted experiments to study use of Containers with two different interconnects (1) InfiniBand for RDMA (2) Ethernet for TCP communication.
% \begin{itemize}
% 	\item InfiniBand for RDMA 
% 	\item Ethernet for TCP communication
% \end{itemize}
In this approach, we considered three classes of MPI benchmarks (1)MiniFE as CPU bound (2)KMI Hash as memory bound and (3) OSU benchmark for latency.
% \begin{itemize}
% 	\item MiniFE as CPU bound 
% 	\item KMI Hash as memory bound 
% 	\item OSU benchmark for latency
% \end{itemize}
%\textcolor{blue}{[Is OSU really bandwidth bound? Check its documentation.]}
For the first two approaches (1) Ethernet on bare metal (Ethernet-BareMetal) and (2) Ethernet on Docker container on host network (Ethernet-Docker), both were used to measure the  performance overhead compared to the third approach, which is bare metal with InfiniBand (InfiniBand-BareMetal). For the experimental evaluation, the average performance of 10 iterations was considered, and the results are presented in Table 4.
\begin{table*}[t]
\centering

\label{my-label}
\begin{tabular}{|c|c|c|c|c|c|}
\hline
\multirow{2}{*}{\textbf{Benchmarks}} & \multirow{2}{*}{\textbf{MPI Ranks}} & \multicolumn{4}{c|}{\textbf{Interconnect Methods}}          				\\ \cline{3-6} 
  &   & \textbf{InfiniBand-BareMetal}          & \textbf{InfiniBand-Docker}        & \textbf{Ethernet-Bare Metal}        & \textbf{Ethernet-Docker}       \\ \hline
MiniFE (MFLOPS)               & 96                         & 94557.7              & 94484.9              & 93612.5              & 93258.6              	\\ \hline
KMI Hash (queries/second)     & 72                         & 4213375.4            & 4209725.1            & 653820.8             & 602075.2             	\\ \hline
OSU Latency (micro second)                  & 96                         & 24074.9              & 24064.7              & 82972.8              & 82994.4               \\ \hline
\end{tabular}
\vspace{0.5em}
\caption{Performance of MPI benchmarks with different interconnect approaches}
\end{table*}

\begin{figure*}%
	\captionsetup[subfigure]{justification=centering}
	\centering
    \subfloat[\it MiniFE - comparing performance over InfiniBand and Ethernet based interconnects for 96 MPI ranks]
    {{\includegraphics[width=0.30\linewidth]
    {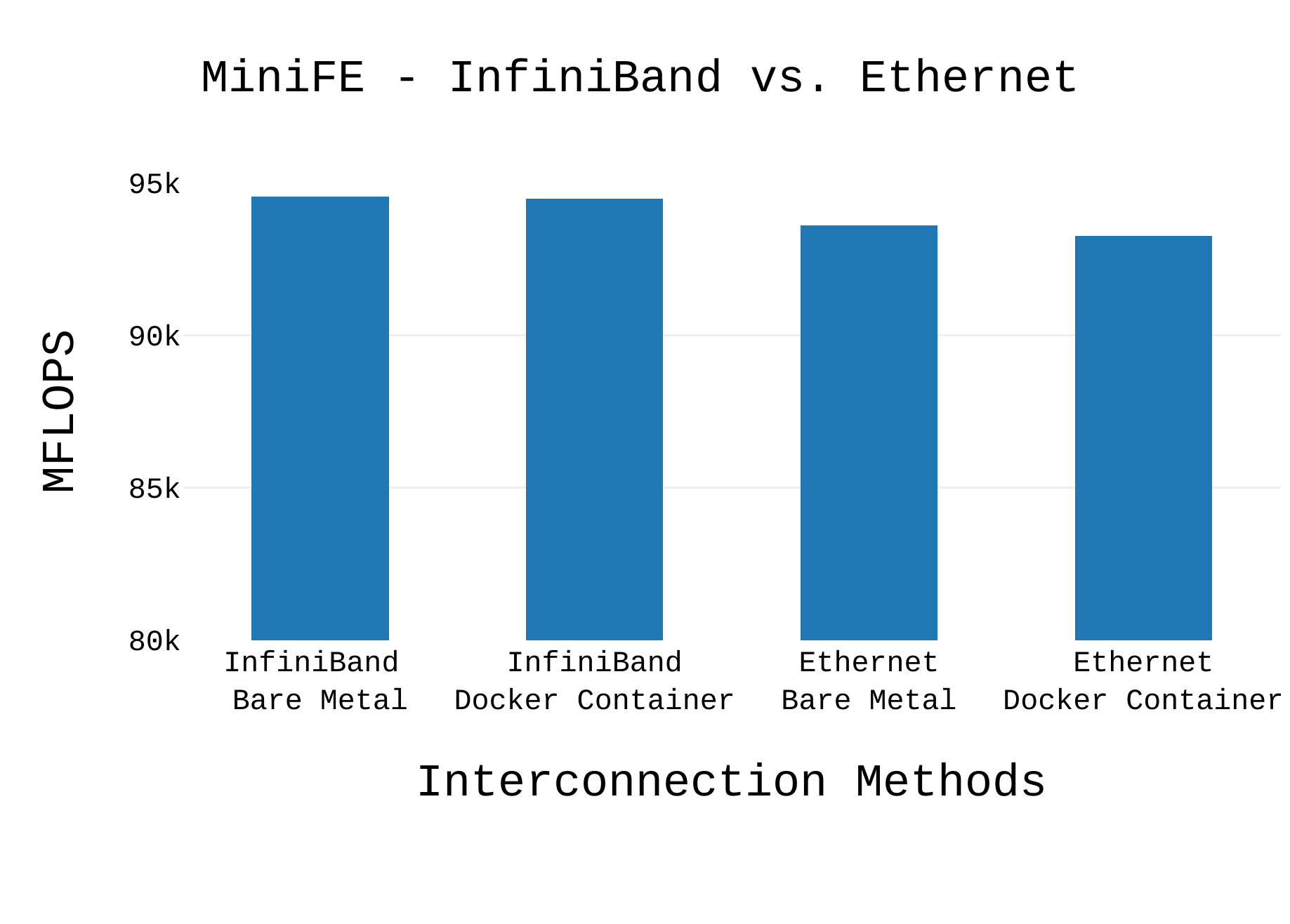}  
    \label{fig:minife_infiniband_ethernet}
    }}
    \subfloat[\it KMI Hash - comparing performance over InfiniBand and Ethernet based interconnects for 72 MPI ranks]
{{\includegraphics[width=0.30\linewidth]
    {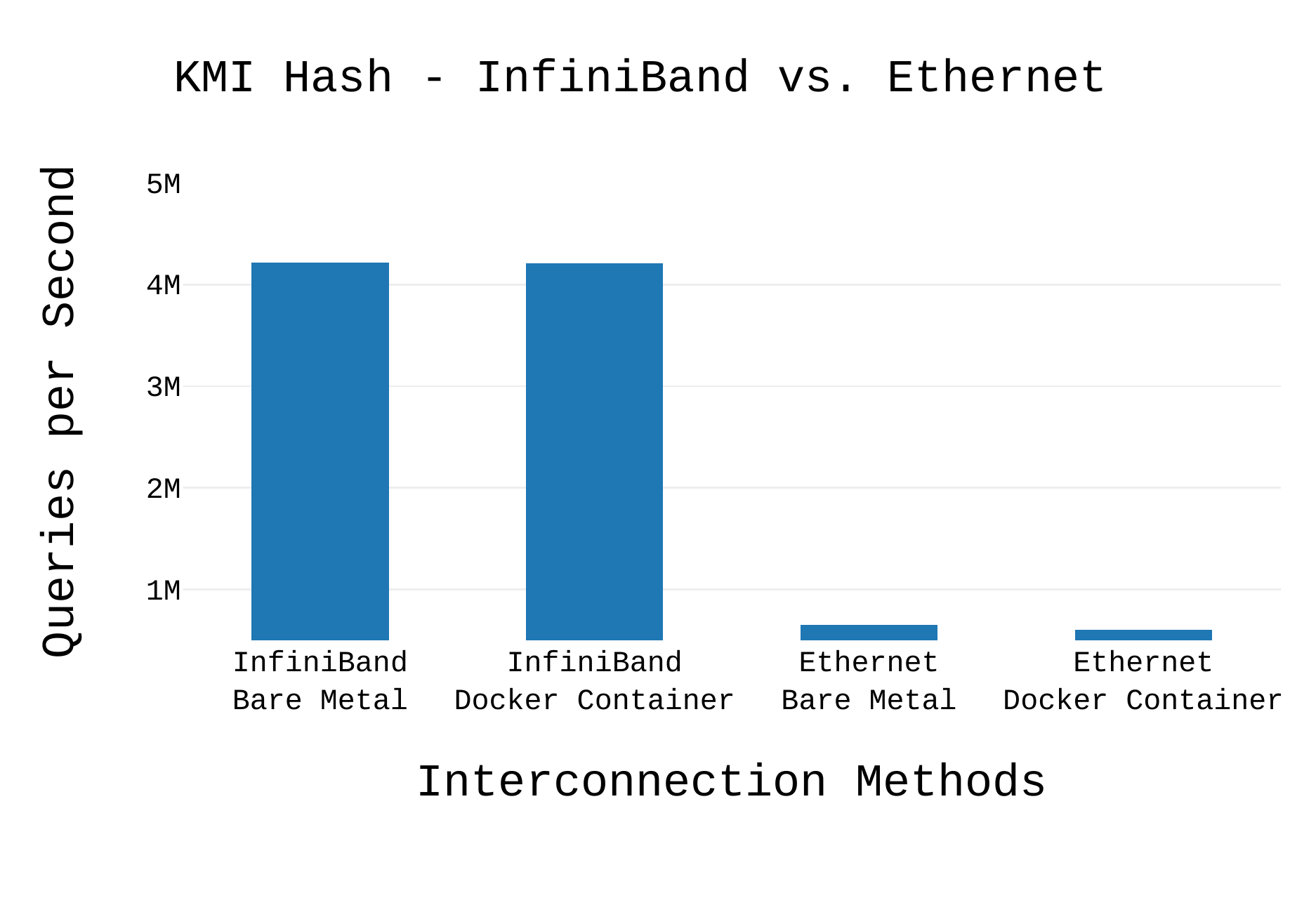}
    \label{fig:hash_infiniband_ethernet}
    }}
    \subfloat[\it OSU alltoallv - comparing performance over InfiniBand and Ethernet based interconnects for 96 MPI ranks]
{{\includegraphics[width=0.30\linewidth]
{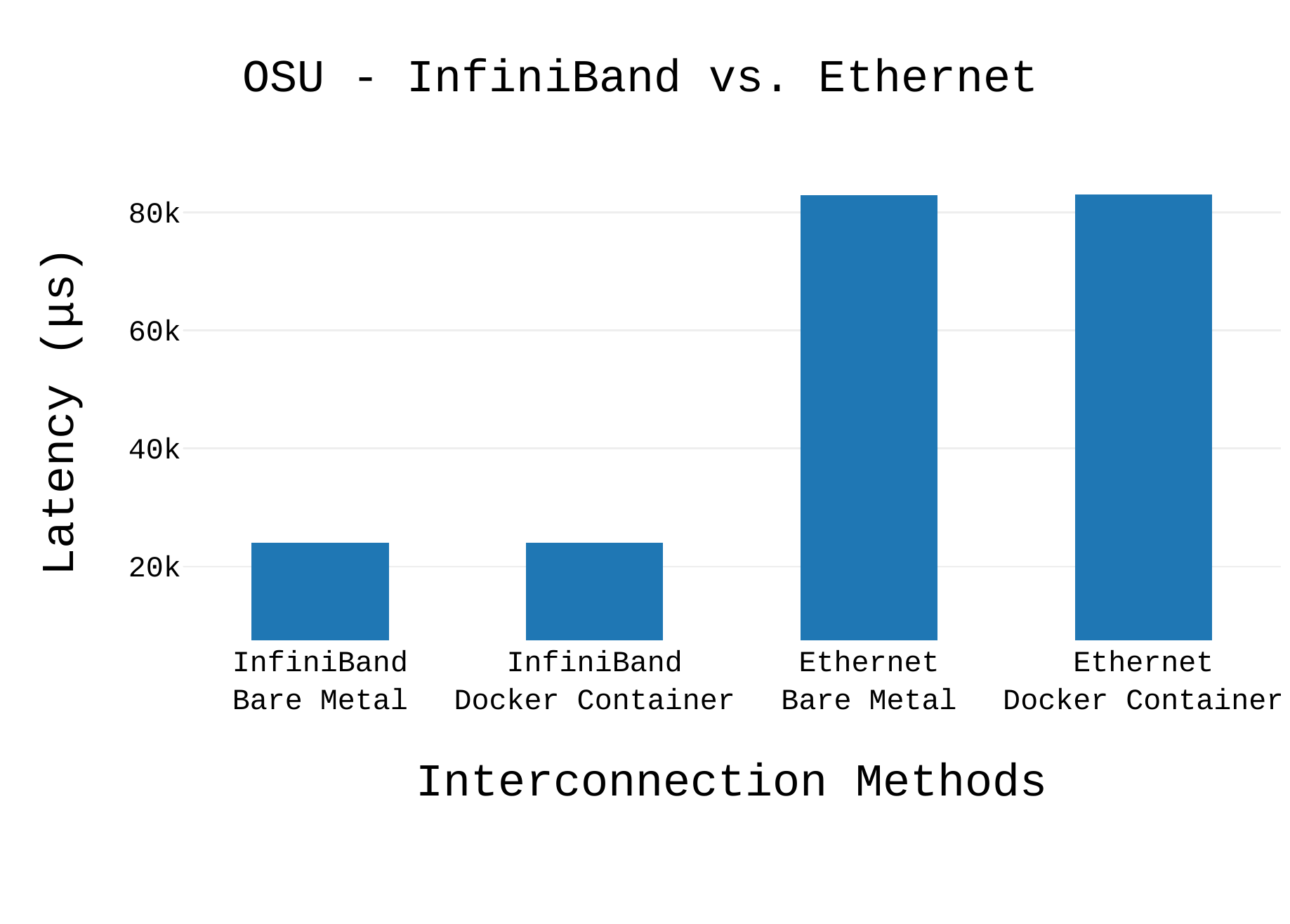}\label{fig:osu_infiniband_ethernet}
    }}
\caption{ Performance of MPI benchmarks with InfinBand and Ethernet interconnect for RDMA and TCP based MPI communication}
\label{fig:infiniEthernet}
\end{figure*}
In Figure~\ref{fig:minife_infiniband_ethernet}, the benchmark MiniFE for Ethernet-BareMetal and Ethernet-Docker approaches performed within 1\% and 1.4\% of the InfiniBand-BareMetal approach respectively. This illustrates that CPU performance hit is minimal. However, Figure~\ref{fig:hash_infiniband_ethernet} shows a similar setup, but this time with a memory intensive benchmark, KMI hash. As expected, when using Ethernet-BareMetal and Ethernet-Docker approaches, we noticed an overhead of 84\% and 85\% respectively. This is a significant performance degradation compared to the InfiniBand-BareMetal approach. Unlike a CPU intensive benchmark, RDMA over InfiniBand interconnects outperforms the TCP based inter-process communication for a memory intensive benchmark such as KMI Hash.
  As expected, the OSU benchmark yielded better latency in the presence of InfiniBand. Figure \ref{fig:osu_infiniband_ethernet} shows that Ethernet-BareMetal and Ethernet-Docker have close to 245\% overhead compared to the InfiniBand-BareMetal approach. As presented in Figure \ref{fig:infiniEthernet}, InfiniBand-Docker performed within  1\% of the  InfiniBand-BareMetal setup. 
  
%We observed that the key performance differences come from using InfiniBand interconnect. It is important to know that for bandwidth and memory bound applications performance can be substantially degraded if the containers present limitations in assessing the host machine's InfiniBand interconnects. Hence, it is essential to invest in container design and deployment.

%% file: 6_related_work.tex
\section{Related Work}
%Enabling Docker Containers for High-Performance and Many-Task Computing.

%Several previous efforts were put together to bring modern day containerization technologies in to High Performance Computing echo systems. 
Docker is the widely used and supported containerization solution in the industry, but its adoption int HPC is hindered due to Docker's root escalation concerns. Azab et al.~\cite{Azab2017EnablingComputing} developed a secure way of running Docker containers in HPC via a Slurm scheduler, without altering the underlying Docker engine and utilizing the full potential of Docker containers.
%\textcolor{blue}{[Need a line on their conclusions.]}
%\textcolor{red}{Such efforts are essential since they can lead to improvements in the HPC community to efficiently run applications in commodity shared hardware.}

%Provision od Docker and InfiniBand in High Performance Computing.
Apart from the security concerns due to root escalation, bandwidth and throughput for HPC jobs via Docker container has been another concern. HPC jobs require faster interconnection across ranks for better performance. 
%InfiniBand (IB) interconnect is a HPC communication standard to overcome the latency due to MPI communication across host nodes.
Unlike Singularity containers, Docker does not support InfiniBand (IB) interconnect as part of its architecture, but Chung et al.~\cite{Chung2016ProvisionComputing} deployed Docker on an IB setup and evaluated the performance of containers over IB with other visualization technologies. Chung et al.'s research also aimed to highlight the benefits of IB with Docker containers. 
%Our focus in in this paper is on HPC technologies, while their experiments for is visualization applications. 

%Scylla: A mesos framework for containerize MPI jobs.
% integrating apache airavata with mesos and docker.

%a tale of two sytems: using containers to deploy hpc application on super computers and cloud.

Younge et al. have defined a model for parallel MPI application DevOps for HPC systems to improve the development effort and reproducibility with the help of containers. They evaluated the feasibility of containers in HPC and showed the performance of Singularity containers on Cray systems~\cite{Younge2017AClouds}.

In our previous work~\cite{Saha2017Scylla:Jobs} \cite{Saha2016IntegratingMesos}, we have shown how Docker containers can be integrated with HPC environments and run MPI applications with cloud-enabled schedulers like Apache Mesos.

%\textcolor{red}{Their researched was orchestrated on vendor-specific environments as well as commercial clouds such as Amazon Elastic Compute Cloud (EC2). while the contributions made by Younge et al. are worth highlighting, the limitations presented by using vendor-specific software libraries and hardware hindered any major conclusions about the feasibility of container technologies in HPC environments.}

%% file: 7_conclusion.tex
\section{Conclusions}
\begin{itemize}

\item Containers can be used to make HPC applications portable. They have proven to provide flexibility and maintainability for commercial applications executing on clouds. 

\item We conducted experiments to determine the performance of different benchmarks on {\it Intel(R) Xeon(R) CPU E5-2670 v3 @ 2.30GHz} based cloud nodes. The performance of different containerization approaches are extremely close to bare metal. Different modes of running MPI application over a private cloud provides both flexibility and minimal performance overhead (less than 1\%).

\item Singularity provides direct support for MPI, and while Docker still does not provide full support for MPI, it is another choice developers and administrator can make. Docker provides more flexibility in terms of container placement with fine-grained resource allocation. 

\item Unlike Singularity, a Docker container needs to have InfiniBand interconnect drivers installed and mapped inside the container to enable fast communication.

\item For MPI applications, splitting ranks per container with restricted resources to each container can be employed by Docker. This option is not available in Singularity containers. 

\end{itemize}

%% file: 8_acknowledgements.tex
\appendix
%Appendix A
\section{Headings in Appendices}
\subsection{Introduction}
\subsection{Background}
\subsubsection{Docker Container}
\subsubsection{Docker Swarm Mode}
\subsubsection{Singularity Containers}
\subsubsection{InfiniBand}

\subsection{Experimental Setup}
\subsubsection{ Bare Metal Nodes + InfiniBand (IB)}
\subsubsection{Docker: one container per node, host
network + InfiniBand}
\subsubsection{Docker: one container per node, overlay
network + InfiniBand}
\subsubsection{Docker - Multiple containers per Host and
\lq n\rq ~MPI Ranks per Container}
\subsubsection{Singularity Container}
\subsection{Evaluation}
\subsubsection{HPCG -High Performance Conjugate
Gradients}
\subsubsection{ MiniFE - Finite Element mini-application}
\subsubsection{OSU - Ohio State University Micro
benchmarks}
\subsubsection{KMI Hash - M-mer Matching Interface
benchmark}
\subsubsection{ Evaluating InfiniBand and Ethernet for
different classes of MPI applications}

\subsection{Related Work}
\subsection{Conclusions}
\subsection{References}

\begin{acks}
This work is partially supported by National Science Foundation, through the OAC-1740263 award.
\end{acks}